\documentclass[fleqn,usenatbib]{mnras}

\usepackage{newtxtext,newtxmath}
\usepackage[T1]{fontenc}

\DeclareRobustCommand{\VAN}[3]{#2}
\let\VANthebibliography\thebibliography
\def\thebibliography{\DeclareRobustCommand{\VAN}[3]{##3}\VANthebibliography}

\usepackage{graphicx}	% Including figure files
\usepackage{amsmath}	% Advanced maths commands
\hypersetup{breaklinks=true}
\usepackage[usenames,dvipsnames]{xcolor}
\usepackage{lineno}
\graphicspath{{./}{figures/}}

\newcommand\editone[1]{\textcolor{black}{#1}}

\title[A CME encountered by four probes inside 1 au]{A coronal mass ejection encountered by four spacecraft within 1~au from the Sun: Ensemble modelling of propagation and magnetic structure}

\author[E. Palmerio et al.]{
Erika~Palmerio,$^{1}$\thanks{E-mail: epalmerio@predsci.com (EP)}
Christina~Kay,$^{2}$
Nada~Al-Haddad,$^{3}$
Benjamin~J.~Lynch,$^{4,5}$
Domenico~Trotta,$^{6}$
\newauthor{
Wenyuan~Yu,$^{3}$
Vincent~E.~Ledvina,$^{7}$
Beatriz~S{\'a}nchez-Cano,$^{8}$
Pete~Riley,$^{1}$
Daniel~Heyner,$^{9}$
}
\newauthor{
Daniel~Schmid,$^{10}$
David~Fischer,$^{10}$
Ingo~Richter,$^{9}$
and Hans-Ulrich Auster$^{9}$
}
\\\\
$^{1}$Predictive Science Inc., San Diego, CA 92121, USA\\
$^{2}$The Johns Hopkins University Applied Physics Laboratory, Laurel, MD 20723, USA\\
$^{3}$Space Science Center, University of New Hampshire, Durham, NH 03824, USA\\
$^{4}$Department of Earth, Planetary, and Space Sciences, University of California, Los Angeles, CA 90095, USA\\
$^{5}$Space Sciences Laboratory, University of California, Berkeley, CA 94720, USA\\
$^{6}$The Blackett Laboratory, Department of Physics, Imperial College London, London, SW7 2AZ, UK\\
$^{7}$Geophysical Institute, University of Alaska Fairbanks, Fairbanks, AK 99775, USA\\
$^{8}$School of Physics and Astronomy, University of Leicester, Leicester, LE1 7RH, UK\\
$^{9}$Institut f{\"u}r Geophysik und extraterrestrische Physik, TU Braunschweig, D-38106 Braunschweig, Germany\\
$^{10}$Space Research Institute, Austrian Academy of Sciences, A-8042 Graz, Austria
}

\date{Accepted XXX. Received YYY; in original form ZZZ}

\pubyear{2024}

\begin{document}
\label{firstpage}
\pagerange{\pageref{firstpage}--\pageref{lastpage}}
\maketitle

%% -------------------------------------- %%
%  ABSTRACT
%% -------------------------------------- %%
 
\begin{abstract}
Understanding and predicting the structure and evolution of coronal mass ejections (CMEs) in the heliosphere remains one of the most sought-after goals in heliophysics and space weather research. A powerful tool for improving current knowledge and capabilities consists of multi-spacecraft observations of the same event, which take place when two or more spacecraft fortuitously find themselves in the path of a single CME. Multi-probe events can not only supply useful data to evaluate the large-scale of CMEs from 1D in-situ trajectories, but also provide additional constraints and validation opportunities for CME propagation models. In this work, we analyse and simulate the coronal and heliospheric evolution of a slow, streamer-blowout CME that erupted on 23 September 2021 and was encountered in situ by four spacecraft approximately equally distributed in heliocentric distance between 0.4 and 1~au. We employ the Open Solar Physics Rapid Ensemble Information (OSPREI) modelling suite in ensemble mode to predict the CME arrival and structure in a hindcast fashion and to compute the ``best-fit'' solutions at the different spacecraft individually and together. We find that the spread in the predicted quantities increases with heliocentric distance, suggesting that there may be a maximum (angular and radial) separation between an inner and an outer probe beyond which estimates of the in-situ magnetic field orientation (parameterised by flux rope model geometry) increasingly diverge. We discuss the importance of these exceptional observations and the results of our investigation in the context of advancing our understanding of CME structure and evolution as well as improving space weather forecasts.
\end{abstract}

% Select between one and six entries from the list of approved keywords.
% Don't make up new ones.

%% -------------------------------------- %%
%  INTRODUCTION
%% -------------------------------------- %%

\section{Introduction} \label{sec:intro}

One of the ultimate goals in heliophysics is to achieve a full characterisation of the structure and evolution of coronal mass ejections (CMEs) from their eruption through their heliospheric propagation. This is important not only from a fundamental physics perspective, but also for space weather science and operations, since CMEs are well-known to generally be the drivers of the most intense geomagnetic storms \citep[e.g.,][]{zhang2007, temmer2021}. The overall picture that has emerged after a few decades of research is that, irrespective of their pre-eruptive configuration \citep[see][and references therein]{patsourakos2020}, CMEs leave the Sun as flux ropes \citep[e.g.,][]{forbes2000, green2018}, which consist of bundles of twisted magnetic fields that warp about a central axis. After a phase of rapid acceleration and expansion in the lower corona \citep[e.g.,][]{patsourakos2010, balmaceda2022} due to their large internal pressure compared to the surrounding environment \citep[e.g.,][]{attrill2007, zhuang2022}, CMEs tend to propagate in a self-similar fashion \citep[e.g.,][]{demoulin2009, subramanian2014} until ${\sim}10$--15~au, when they reach pressure balance with the ambient solar wind \citep[e.g.,][]{richardson2006, vonsteiger2006}. However, the specific evolution of a given CME may deviate substantially from this idealised scenario due to a multitude of possible factors \citep[e.g.,][]{manchester2017, luhmann2020}.

In most cases, in fact, CMEs do not propagate through a uniform background, but through a structured medium that consists of different solar wind flows \citep[e.g.,][]{maunder2022, palmerio2022b}, slow--fast stream interaction regions \citep[e.g.,][]{wang2014, alshakarchi2018}, the heliospheric current/plasma sheet \citep[e.g.,][]{blanco2011, liu2016}, and even other CMEs \citep[e.g.,][]{lugaz2017, trotta2024b}. Outcomes of these interaction processes include deflections \citep[e.g.,][]{lugaz2012, zuccarello2012}, rotations \citep[e.g.,][]{vourlidas2011, liu2018}, deformations \citep[e.g.,][]{liu2006, savani2010}, and erosion \citep[e.g.,][]{ruffenach2015, pal2020}. As a result, the magnetic configuration of an erupting flux rope at the Sun that is inferred from remote-sensing observations \citep[see][and references therein]{palmerio2017} may differ more or less dramatically from the one that is then measured in situ \citep[e.g.,][]{yurchyshyn2007, palmerio2018, xie2021}. More so, the specific trajectory through a given CME that is sampled by a spacecraft may not even be representative of the structure as a whole because of local distortions \citep[e.g.,][]{owens2020}. To complicate things further, analyses of the in-situ structure of CMEs are often performed with the aid of flux rope fitting/reconstruction models, each based on a certain geometry and physical assumption. However, studies have shown that different flux rope fitting technique can provide very different results for the same CME \citep[e.g.,][]{riley2004, alhaddad2013}, albeit it appears that the level of agreement across models increases for ``simpler'' CMEs that display little signatures of expansion and generally more symmetric magnetic field profiles \citep[e.g.,][]{alhaddad2018}.

The complexity of the myriad processes dictating CME evolution in interplanetary space, together with the known limitations of the available analysis techniques, make it clear that determining the global configuration of a CME from single-spacecraft measurements is a particularly arduous task. For this reason, a number of studies have attempted to obtain a more complete insight into CME structure and evolution using fortuitous relative configurations of two or more probes that have detected the same event in situ. A notable example is the work of \citet{burlaga1981}, who reported observations of a single CME in January 1978 by five spacecraft that were distributed over ${\sim}30^{\circ}$ in longitude between ${\sim}1$--2~au from the Sun. In fact, this was the first study to report that all observing probes detected a ``magnetic loop'' structure that is now known as a magnetic cloud, i.e.\ an ejecta that is characterised by enhanced magnetic field strength, smoothly rotating magnetic field vectors, declining speed profiles, as well as depressed temperature and plasma beta---generally interpreted as the in-situ signatures of a flux rope. The potential of multi-spacecraft observations has gained significant traction over recent years, so much so that it is possible nowadays to find a few dedicated catalogues in the existing literature \citep[e.g.,][]{davies2022, mostl2022}. However, most multi-probe encounters are realised over arbitrary spatial separations of the observers involved, making it difficult to attribute e.g.\ structural differences to temporal evolution, to local distortions, or to both \citep[e.g.,][]{riley2003, dumbovic2019}. Additionally, some events are measured over very large (i.e., of at least a few au) radial separations between the observing probes, in which case not even a near-longitudinal alignment would grant that the exact same structure has been detected due to repeated interactions with the structured background \citep[e.g.,][]{burlaga2001, palmerio2021c}. Nevertheless, some studies have attempted to isolate these processes by focussing on events characterised by spacecraft close to radial alignment \citep[e.g.,][]{good2019, vrsnak2019, salman2020, winslow2021}, or with probes spread in longitude at the same radial distance \citep[e.g.,][]{kilpua2009, farrugia2011, lugaz2022, carcaboso2024}, or where observations from two relatively nearby locations are available \citep[e.g.,][]{lugaz2018, davies2021, palmerio2024b, regnault2024}.

As an additional indication that a deep knowledge of CME structure and evolution in the heliosphere is still to be achieved, it is worth remarking that multipoint space weather forecasts of CMEs---or, in the case of heliophysics research, most often hindcasts---have been centred largely on arrival times and/or arrival speeds at multiple locations \citep[e.g.,][]{witasse2017, palmerio2021a} rather than on the magnetic field configurations upon impact \citep[e.g.,][]{asvestari2021, sarkar2024}. Truth to be told, this is usually also the case for single-spacecraft encounters \citep[e.g.,][]{riley2018, kay2024b} given the well-known challenges associated with magnetic fields forecasts \citep[e.g.,][]{kilpua2019}, but it is only natural to assume that difficulties (and uncertainties) in predicting CME magnetic structure can only increase with the number of observers available for model--data comparisons. On the other hand, the power of multi-probe events is exactly that they allow for models to be validated not at a single location (e.g., Earth), but throughout a specific interval of a CME's journey away from the Sun, permitting thus to increase our understanding of how it evolves and/or of how its local structure compares to the global one.

%% -------------------------------------- %%
%% Figure: Spacecraft positions
\begin{figure}
\centering
\includegraphics[width=.99\linewidth]{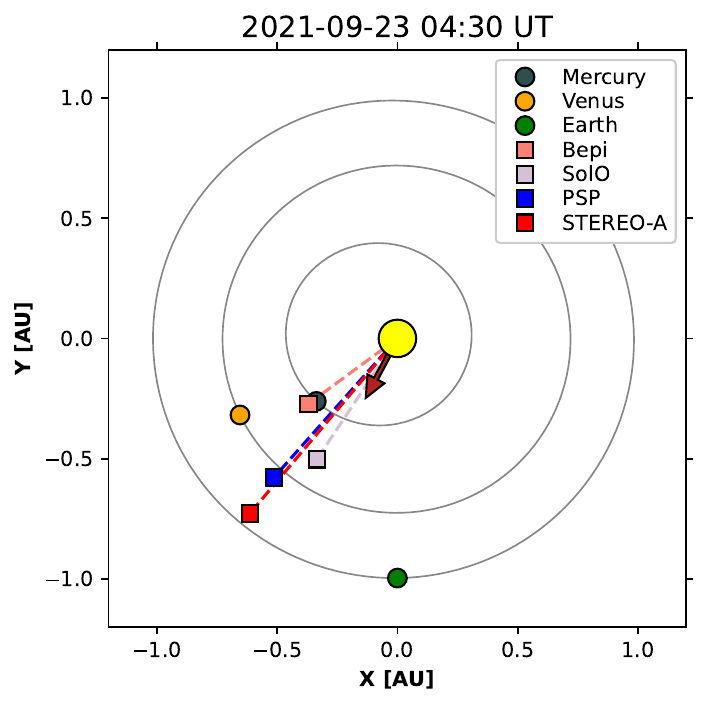}
\caption{Position of planets and spacecraft within 1~au from the Sun on 23 September 2021 at 04:30~UT, i.e.\ around the CME eruption time. The longitude of the CME source region is indicated with an arrow emanating from the surface of the Sun. The four probes that encountered the event under study are connected to the centre of the Sun via dashed lines. The orbits of Mercury, Venus, and Earth are also shown.}
\label{fig:orbits}
\end{figure}
%% -------------------------------------- %%

In this work, we analyse in detail the inner heliospheric evolution of a CME that erupted on 23 September 2021. The remarkable nature of this event resides in the fact that it was detected in situ by four spacecraft that were close to radial alignment and more or less uniformly spread between 0.4 and 1~au (see Figure~\ref{fig:orbits}), namely BepiColombo \citep[Bepi;][]{benkhoff2021}, Solar Orbiter \citep[SolO;][]{muller2020}, Parker Solar Probe \citep[PSP;][]{fox2016}, and Solar Terrestrial Relations Observatory Ahead \citep[STEREO-A;][]{kaiser2008}. Our aim in studying this fortuitous encounter is to closely follow its evolution from the Sun through the inner heliosphere and to perform a multi-probe hindcast of its structure---i.e., testing how well our models can reproduce the propagation of CME flux ropes at different points in space and time. This manuscript is structured as follows. In Section~\ref{sec:remote}, we provide an overview of the remote-sensing observations associated with the 23 September 2021 eruption. In Section~\ref{sec:insitu}, we present and analyse the interplanetary measurements of the CME under study at the four different observers. In Section~\ref{sec:osprei}, we perform hindcasts of the event using the Open Solar Physics Rapid Ensemble Information \citep[OSPREI;][]{kay2022a} analytical modelling suite, with particular emphasis on the CME magnetic structure. In Section~\ref{sec:discussion}, we discuss the 23 September 2021 event from both an observational and a modelling perspective. Finally, in Section~\ref{sec:concl}, we summarise our findings and draw our conclusions.

%% -------------------------------------- %%
%  SOLAR OBSERVATIONS
%% -------------------------------------- %%

\section{Overview of the solar observations} \label{sec:remote}

The eruption and subsequent coronal propagation of the 23 September 2021 CME analysed in this work were observed in extreme ultra-violet (EUV) and white-light (WL) imagery from two viewpoints, i.e.\ Earth and the STEREO-A spacecraft. For Earth's perspective, we use solar disc imagery from the Atmospheric Imaging Assembly \citep[AIA;][]{lemen2012} onboard the Solar Dynamics Observatory \citep[SDO;][]{pesnell2012} as well as coronagraph data from the C2 and C3 cameras part of the Large Angle and Spectrometric Coronagraph \citep[LASCO;][]{brueckner1995} onboard the Solar and Heliospheric Observatory \citep[SOHO;][]{domingo1995}. From STEREO-A, we employ images of the solar disc from the Extreme Ultraviolet Imager \citep[EUVI;][]{wuelser2004} and coronagraph data from the COR2 camera, both part of the Sun Earth Connection Coronal and Heliospheric Investigation \citep[SECCHI;][]{howard2008} suite. Additionally, we take advantage of magnetograph data collected by the Helioseismic and Magnetic Imager \citep[HMI;][]{scherrer2012} onboard SDO.

\subsection{Source region and eruption} \label{subsec:sun}

An overview of the available EUV observations for this event is shown in Figure~\ref{fig:obs_euv} and a full-disc animated version is provided in Supplementary Video~1. The Sun appears rather active around the time of interest, with several eruptions lifting off the visible disc as well as the limb from both the Earth and STEREO-A perspectives. The eruptive event that is the main focus of this work originates from NOAA active region (AR) 12871 on 23 September 2021 around 04:30~UT, and is accompanied by an M2.8 flare peaking at 04:42~UT. In STEREO-A imagery (Figure~\ref{fig:obs_euv}(a--b)), the CME source region is located on the southwestern quadrant and the sequence of events features the eruption itself from the western portion of AR~12871 (on-disc arrow in Figure~\ref{fig:obs_euv}(a)), the lift-off of a large double-loop structure (off-limb arrows in Figure~\ref{fig:obs_euv}(a)), and the subsequent appearance of an additional set of post-eruption arcades (PEAs) in the eastern portion of AR~12871 (arrow in Figure~\ref{fig:obs_euv}(b)). These complex observations can be further interpreted using imagery from Earth orbit, where magnetograph measurements complement the EUV data. From the SDO viewpoint (Figure~\ref{fig:obs_euv}(c--d)), the CME source region is located on the southeastern quadrant and on-disc signatures of the eruption include loops opening and propagating northwards of AR~12871 (see Supplementary Video~1, marked with a yellow dashed curve in Figure~\ref{fig:obs_euv}(c)) in addition to the PEA systems identified in STEREO-A imagery (arrows in Figure~\ref{fig:obs_euv}(c--d)).

The magnetogram contours shown over the EUV data in Figure~\ref{fig:obs_euv}(c) reveal a complex structure of the local photosphere, with different polarity patches forming nested flux systems---distinct regions of closed flux embedded within the larger-scale flux system of a coronal streamer \citep[e.g.,][]{longcope2005, karpen2024}. Such a configuration tends to result in curved polarity inversion lines (PILs), with a separatrix dome at the interface between different polarities \citep[e.g.,][]{wyper2016a, wyper2016b}, as is evident from the PIL contours displayed in Figure~\ref{fig:obs_euv}(d). This is also a structure that often generates circular-ribbon flares \citep[e.g.,][]{lee2020, zhang2024}. The closed-field topology of AR~12871 lies beneath a helmet streamer that is largely east--west oriented, curves around the approximately C-shaped PIL, and continues back toward the west. 

%% -------------------------------------- %%
%% Figure: Solar disc observations
\begin{figure}
\centering
\includegraphics[width=.99\linewidth]{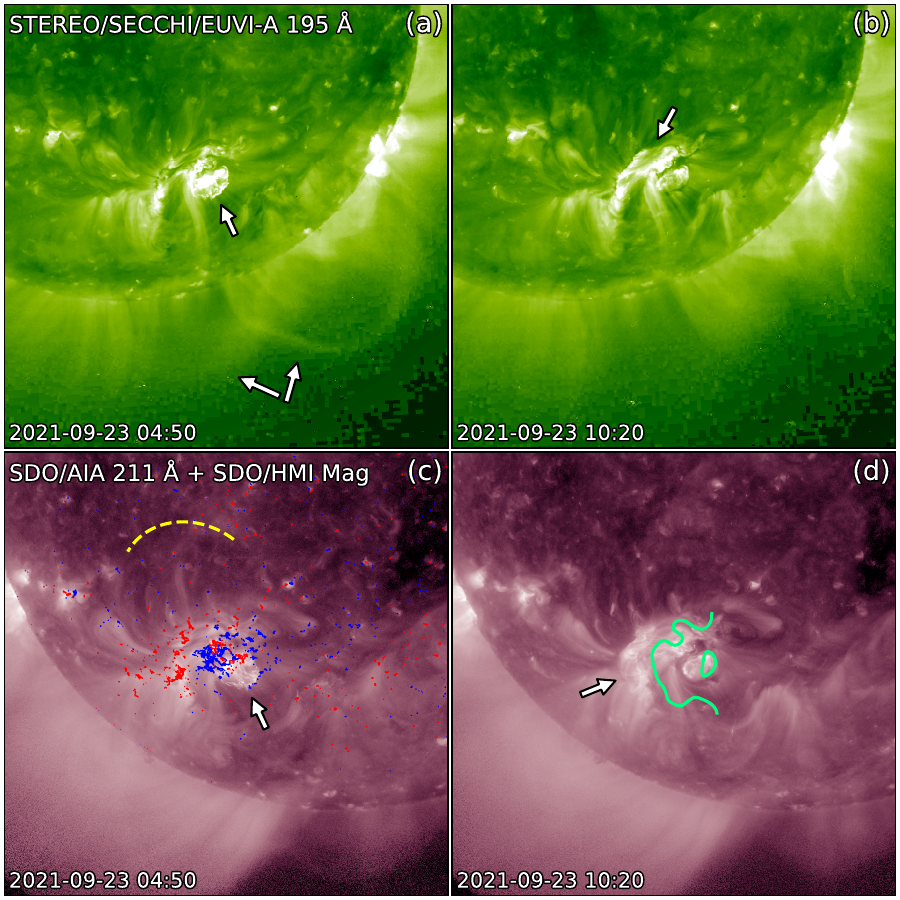}
\caption{Overview of some EUV observations available for the 2021 September 23 eruption, from the (a--b) STEREO-A and (c--d) SDO (Earth) perspectives. In the bottom panels, EUV observations are complemented by (c) magnetogram contours saturated at \editone{${\pm}150$~G} (red: positive polarity; blue: negative polarity), and (d) the polarity inversion line associated with AR 12871 (green contour), obtained from smoothed magnetograph data. Throughout the panels, the various arrows and the dashed yellow line highlight interesting features associated with the eruption (see the main text for details).}
\label{fig:obs_euv}
\end{figure}
%% -------------------------------------- %%

In Supplementary Video~1, the first set of PEAs appears at 04:40--04:45~UT with a simultaneous remote brightening at the footpoint of the external spine line \citep[reminiscent of the ``EIT crinkles'' of][]{sterling2001}, followed by a clear dimming in AIA 211~{\AA} imagery surrounding the dome and a whole loop-shaped region along the external spine flux tube connecting the dome and the remote brightening area. This coincides exactly with the eruption and expansion of the overlapping off-limb loops in the EUVI 195~{\AA} data. The apparent overlap of these loops results from emission of different structures in the optically thin corona being summed during the line-of-sight integration, as these are two different parts of the same streamer belt flux system making the U-turn. The second, more southern (in projection) loop is more diffuse in emission, but expands and erupts essentially in tandem with the more resolved, westward loop. The AIA 211~{\AA} on-disc dimmings follow the EUVI 195~{\AA} off-limb erupting loops, suggesting that the westward loop portion of the helmet streamer is approximately ``above'' the circular-ribbon flare whilst the southern loop portion is approximately ``above'' the second set of PEAs along the northern half of the C-shaped PIL. The external spine loop from the initial eruption starts brightening again after $\sim$07:00~UT \citep[as in][]{lee2020}, which is followed thereafter by the dimming areas above and below the second PEA gradually returning to their original intensities. Regardless of the specific form of the large-scale helmet streamer energisation, a sufficient expansion (gradual or rapid) of the middle-to-outer layers of the streamer belt closed flux system has been shown to erupt as a streamer-blowout CME \citep[][and references therein]{lynch2016}. 

\subsection{Coronal evolution} \label{subsec:corona}

An overview of the available WL observations for this event is shown in Figure~\ref{fig:obs_wl} and an animated version is provided in Supplementary Video~2. In imagery from both viewpoints (i.e., STEREO-A and SOHO), it is clear that multiple faint eruptions are concurrently present at any time over the period of interest, making interpretation of the different structures (and their origin) especially challenging. In particular, the sequence of events evident in coronagraph data include (in relation to the STEREO-A viewpoint): (1) a streamer blowout originating from near the southeastern limb \editone{and emerging around 06:00~UT}, (2) a jet-like CME associated with the first set of PEAs described in Section~\ref{subsec:sun} and propagating towards the southwest \editone{also around 06:00~UT}, (3) a large-scale streamer blowout associated with the second set of PEAs described in Section~\ref{subsec:sun} and appearing as a (partial) halo \editone{starting around 08:00~UT}, and (4) an additional jet-like CME related to a later (${\sim}$15:30~UT) eruption from AR~12871 \editone{visible from around 16:30~UT}. Hence, the eruptive event that is the focus of this work is the result of a multi-stage nested-flux system eruption of the \citet{karpen2024} type, where the first dome-shaped PEA and remote brightening creates a significant enough disturbance in the streamer flux system that some previously-closed flux opens and that erupting plasma makes it into the open field and solar wind. This jet-like transient triggers a more traditional streamer-blowout eruption above the adjacent PIL leading to the partial halo CME \citep[see][for another example of a jet destabilising an energised, multipolar flux system]{pal2022}. Since both jets appear rather narrow as well as southwards-directed and the first streamer blowout propagates mainly off the eastern limb \editone{(its corresponding solar eruption can be observed in Supplementary Video~1 off the southeastern quadrant around 01:00~UT from STEREO-A's perspective)}, in the following we shall focus on the second streamer blowout, i.e.\ the eruption associated with the second set of PEAs described in Section~\ref{subsec:sun} and that is expected to be encountered in situ by the four nearly-aligned spacecraft shown in Figure~\ref{fig:orbits}.

%% -------------------------------------- %%
%% Figure: Coronagraph observations
\begin{figure}
\centering
\includegraphics[width=.99\linewidth]{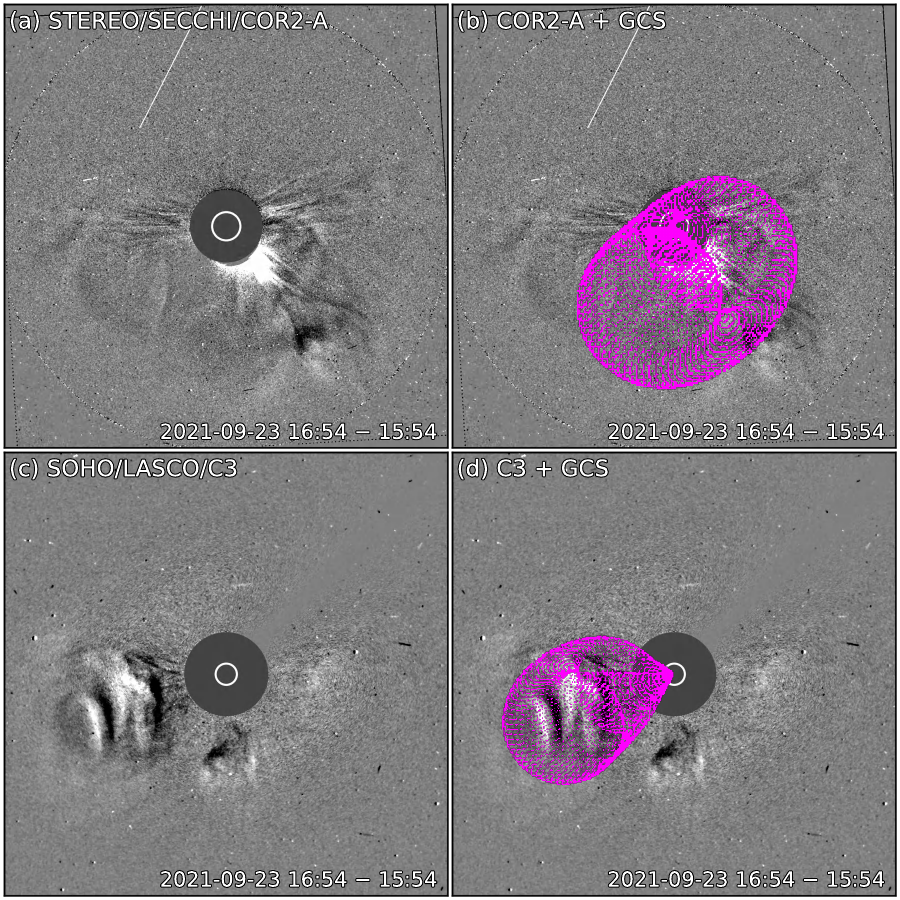}
\caption{The faint 23 September 2021 CME observed in WL imagery from (top) STEREO-A and (bottom) SOHO. The left panels show plain coronagraph difference images, whilst the right panels display the same data with the GCS wireframe overlaid.}
\label{fig:obs_wl}
\end{figure}
%% -------------------------------------- %%

To obtain a first-order assessment of the CME morphology and kinematics through the solar corona, we apply the Graduated Cylindrical Shell \citep[GCS;][]{thernisien2011} model to nearly-simultaneous imagery from STEREO-A and SOHO. The technique consists of manually fitting a parameterised shell (with six free parameters) onto coronagraph imagery, and results at one sample time are shown in the right panels of Figure~\ref{fig:obs_wl}. According to the performed reconstructions, the CME propagates in the direction ($\theta$, $\phi$) = ($-8^{\circ}$, $-37^{\circ}$) in Stonyhurst coordinates with a moderate inclination ($40^{\circ}$ counterclockwise from solar west) to the solar equatorial plane and a slow (${\sim}400$~km$\cdot$s$^{-1}$) speed, as is often the case for streamer blowouts \citep[e.g.,][]{vourlidas2018}. We remark that, apart from uncertainties intrinsic to coronal reconstruction methods performed ``by eye'' \citep[e.g.,][]{verbeke2023, kay2024a}, this specific event is characterised by additional ambiguities due to both its faint nature and its overlapping with other eruptions in projected plane-of-sky images. In fact, it is especially difficult to clearly distinguish the fronts of the two streamer blowouts mentioned above, the second of which is our CME of interest. We do not exclude that the two eruptions may have interacted via their flanks, but since their nose trajectories differ by ${\sim}35^{\circ}$ in longitude \editone{(determined after performing separate GCS reconstructions of both structures)}, it is expected that the in-situ encounters presented in the next section were realised for the most part with the partial-halo CME.

%% -------------------------------------- %%
%  IN-SITU OBSERVATIONS
%% -------------------------------------- %%

\section{Analysis of the interplanetary data} \label{sec:insitu}

\begin{figure*}
\centering
\includegraphics[width=.95\linewidth]{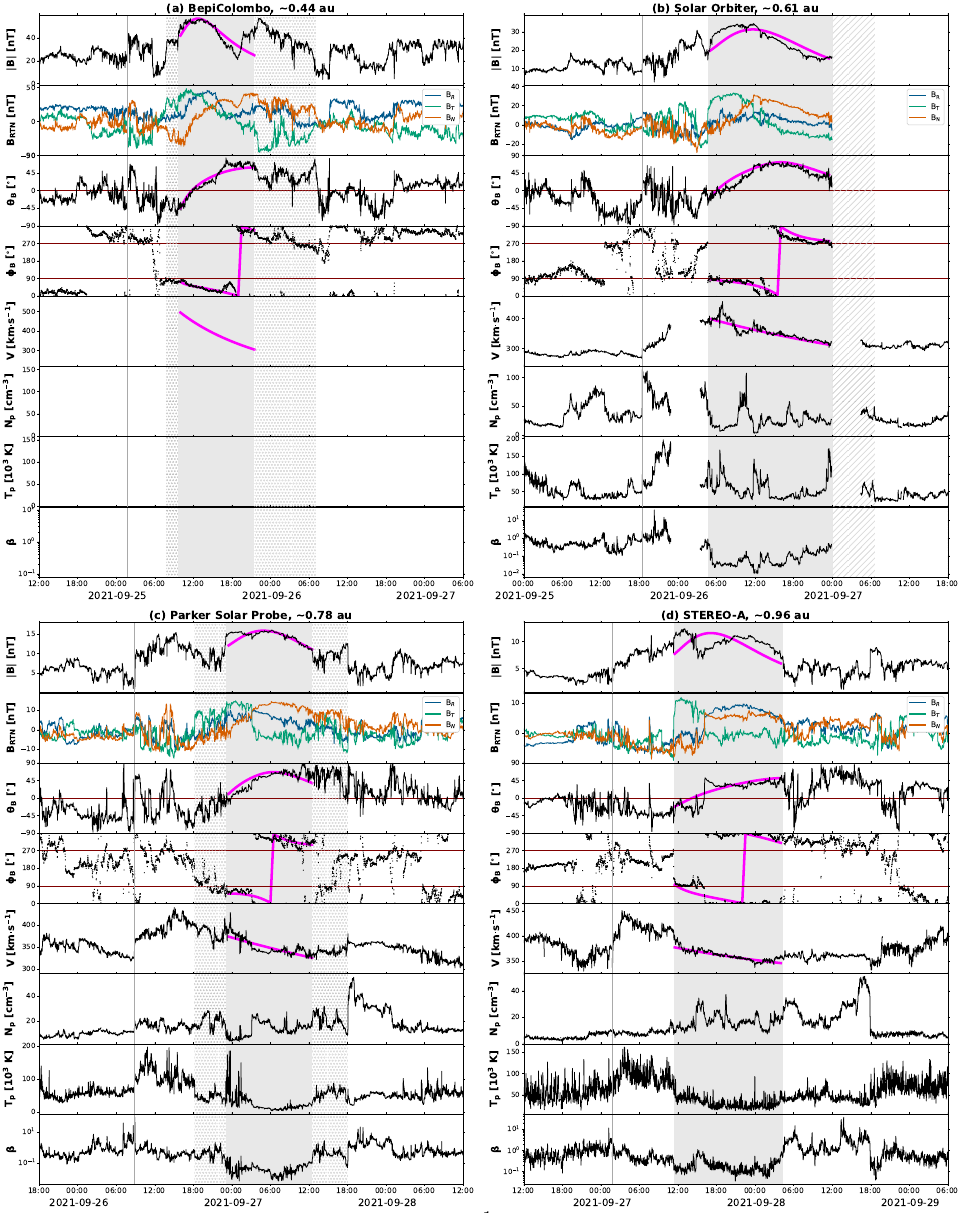}
\caption{In-situ measurements of the 23 September 2021 CME at (a) Bepi, (b) SolO, (c) PSP, and (d) STEREO-A. Each plot shows, from top to bottom: magnetic field magnitude, magnetic field components in Radial--Tangential--Normal (RTN) coordinates, latitudinal and longitudinal angles of the magnetic field, solar wind bulk speed, proton density and temperature, and plasma beta. Interplanetary shocks arrivals are marked with vertical areas, whilst CME ejecta regions are highlighted with shading---the core flux rope in solid colour and other ejecta boundaries in \editone{dotted (Bepi and PSP) or} hatched \editone{(SolO)} markings\editone{---see the main text for details}. The magenta curves show flux rope fitting results using the EFF model. The heliocentric distances reported on top of each plot refer to the respective shock arrival time at each spacecraft (see also Table~\ref{tab:positions}). Note that each panel displays 2.75~days of data.}
\label{fig:obs_insitu}
\end{figure*}

Here, we analyse in detail the magnetic field and plasma measurements of the 23 September 2021 CME at the four probes of interest, i.e.\ Bepi, SolO, PSP, and STEREO-A. An overview of the in-situ measurements collected by the four spacecraft is shown in Figure~\ref{fig:obs_insitu}, and their evolving heliospheric coordinates at the eruption time and as the CME-driven shock impacted each observer are reported in Table~\ref{tab:positions}. Additionally, to evaluate the overall large-scale evolution of the CME through the inner heliosphere and to confirm that the same eruption likely impacted all the four targets, we have performed a magnetohydrodynamic (MHD) simulation using the coupled Wang--Sheeley--Arge \citep[WSA;][]{arge2004} Enlil \citep{odstrcil2003} model---these results are summarised in Appendix~\ref{app:enlil}.

%% -------------------------------------- %%
%% Table: s/c positions in time
\begin{table*}
	\centering
	\caption{Stonyhurst heliographic coordinates in terms of [$r$, $\theta$, $\phi$] triplets (units of [au, deg, deg]) of the four spacecraft at the time of the eruption of the 23 September 2021 CME as well as at the times of the interplanetary shock arrival at each observer.}
	\label{tab:positions}
	\begin{tabular}{l@{\hskip .5in}l@{\hskip .3in}l@{\hskip .3in}l@{\hskip .3in}l@{\hskip .3in}l} 
		\hline
		 & Eruption & Shock at Bepi & Shock at SolO & Shock at PSP & Shock at ST-A\\
		 & 2021-09-23T04:30 & 2021-09-25T01:46 & 2021-09-25T18:25 & 2021-09-26T08:50 & 2021-09-27T01:51\\
		\hline
		Bepi & [0.46, $-$0.3, $-$53.7] & [0.44, $-$0.1, $-$49.3] & [0.44, $-$0.1, $-$47.6] & [0.43, $-$0.0, $-$45.9] & [0.42, $+$0.1, $-$44.1]\\
		SolO & [0.60, $+$1.7, $-$33.7] & [0.61, $+$1.8, $-$31.1] & [0.61, $+$1.9, $-$30.1] & [0.61, $+$1.9, $-$29.2] & [0.62, $+$2.0, $-$28.4]\\
		PSP & [0.77, $+$3.5, $-$41.6] & [0.78, $+$3.5, $-$42.3] & [0.78, $+$3.5, $-$42.6] & [0.78, $+$3.5, $-$42.9] & [0.78, $+$3.5, $-$43.1]\\
		ST-A & [0.96, $+$6.7, $-$40.1] & [0.96, $+$6.8, $-$40.0] & [0.96, $+$6.8, $-$39.9] & [0.96, $+$6.8, $-$39.9] & [0.96, $+$6.9, $-$39.8]\\
		\hline
	\end{tabular}
\end{table*}
%% -------------------------------------- %%

%% -------------------------------------- %%
%% Figure: In-situ observations
%% -------------------------------------- %%

From each set of spacecraft measurements, we search for various in-situ CME signatures \citep[e.g.,][]{zurbuchen2006, kilpua2017} and identify the passage times of the interplanetary shock, sheath region, and CME ejecta. We also determine the boundaries of the ``core'' flux rope, i.e.\ the period in the in-situ time series characterised by clear magnetic cloud signatures such as smoothly rotating magnetic field vectors and low plasma beta---and that may or may not coincide with the extent of the CME ejecta as a whole \citep[e.g.,][]{richardson2010, kilpua2013}.

We also perform a local shock parameter estimation analysis for all the spacecraft crossings, computing the shock normal ($\hat{\mathrm{n}}_{\mathrm{RTN}}$), shock normal angle ($\theta_{Bn}$), magnetic and gas compression ratios (${r}_B$ and ${r}_n$, respectively), shock speed ($V_{\mathrm{sh}}$), as well as the fast magnetosonic and Alfv{\'e}nic Mach numbers ($M_{\rm{fms}}$ and $M_{A}$, respectively). The latter two are defined as the ratio between the shock speed and the upstream fast magnetosonic and Alfv{\'e}n speeds, respectively. When both magnetic field and plasma data are available, we compute the shock normal with the Mixed Mode Method 3~\citep[MX3;][]{abraham-shrauner1976}, and the shock speed along the shock normal and in the spacecraft reference frame with the mass--flux conservation. When the magnetic field information is available, the shock normal angle is computed using the magnetic coplanarity theorem \citep[MCT;][]{colburn1966}. For a detailed description of these techniques, we refer the reader to \citet{paschmann2000}. For the averaging operation involved in the shock parameters, we use a collection of upstream/downstream windows that we vary systematically from 1 to 8 minutes using the SerPyShock code \citep{trotta2022}. Results are summarised in Table~\ref{tab:shocks}.

%% -------------------------------------- %%
%% Table: Shock parameters
\begin{table}
	\centering
	\caption{Shock parameters measured at each spacecraft. The parameters shown are: shock normal vector ($\hat{\mathrm{n}}_{\mathrm{RTN}}$), shock angle ($\theta_{Bn}$), magnetic compression ratio (${r}_B$), gas compression ratio (${r}_n$), shock speed ($V_{\mathrm{sh}}$), as well as fast magnetosonic and Alfv\'enic Mach numbers ($M_{\rm{fms}}$ and $M_{A}$, respectively). The shock normals are shown in the RTN frame of reference, with $\theta_{Bn}$ expressed in degrees. The shock speed $v_{\mathrm{sh}}$ is expressed in km$\cdot$s$^{-1}$ and it is aligned with the shock normal.}
	\label{tab:shocks}
	\begin{tabular}{l@{\hspace{1.7\tabcolsep}}c@{\hspace{1.7\tabcolsep}}c@{\hspace{1.7\tabcolsep}}c@{\hspace{1.7\tabcolsep}}c@{\hspace{1.7\tabcolsep}}c@{\hspace{1.7\tabcolsep}}c@{\hspace{1.7\tabcolsep}}c}
		\hline
		{} & $\hat{\mathrm{n}}_{\mathrm{RTN}}$ & $\theta_{Bn}$ & ${r}_B$ & ${r}_n$ & $V_{\rm{sh}}$ & $M_{\rm{fms}}$ & $M_{A}$\\
		\hline
		Bepi &[0.82, $-$0.36, $-$0.44]&18& 1.7 & --- & ---  & --- & --- \\
        SolO &[0.88, 0.10, $-$0.47] &27& 1.5 & 3.3 & 299  & 1.7 & 1.5 \\
        PSP  &[0.91, 0.34, $-$0.23]&71& 1.8 & 3.0 & 347 & 2.0  & 1.9 \\
        ST-A &[0.50, 0.83, $-$0.25] &13& 2.8 & --- & --- & ---  & --- \\
		\hline
	\end{tabular}
\end{table}
%% -------------------------------------- %%

Additionally, at each spacecraft, we perform a fit of the corresponding core flux rope interval using an expansion-modified force-free \citep[EFF; e.g.,][]{farrugia1993, yu2022} model, which takes the classic constant-$\alpha$ force-free solution \citep[e.g.,][]{burlaga1988, lepping1990} and adds a time-scale parameter that describes the self-similar CME expansion rate ($\tau_\mathrm{exp}$). The remaining quantities retrieved by the fitting method are axis orientation ($\Theta_{0}$, $\Phi_{0}$), field magnitude along the axis ($B_{0}$), flux rope chirality (\editone{or handedness,} $H$), and impact parameter ($p_{0}$, \editone{i.e., the crossing distance to the symmetry axis normalised by the cross-sectional radius}). Since CME expansion rate can be indirectly estimated from the speed profile \citep[e.g.,][]{owens2005, nieves-chinchilla2018a}, in addition to the magnetic field components we include, where possible, the solar wind speed amongst the flux rope proprieties constraining the fit. The flux rope fitting results at the four probes are reported in Table~\ref{tab:frfits}. In the remainder of this section, we describe in detail observations and analysis of the 23 September 2021 CME in-situ measurements at each impacted spacecraft.

%% -------------------------------------- %%
%% Table: Flux rope fits
\begin{table}
	\centering
	\caption{Flux rope fitting results at each spacecraft. The parameters shown are: latitudinal ($\Theta_{0}$) and longitudinal ($\Phi_{0}$) directions of the flux rope axis, axial magnetic field magnitude ($B_{0}$), chirality ($H$), normalised impact parameter over the flux rope's radius ($p_{0}$), expansion time ($\tau_\mathrm{exp}$), and normalised goodness-of-fit measure (chi-squared) over the magnetic field components ($\chi^{2}_\mathrm{dir}$) as well as magnitude ($\chi^{2}_\mathrm{mag}$).}
	\label{tab:frfits}
	\begin{tabular}{l@{\hspace{2.0\tabcolsep}}c@{\hspace{2.0\tabcolsep}}c@{\hspace{2.0\tabcolsep}}c@{\hspace{2.0\tabcolsep}}c@{\hspace{2.0\tabcolsep}}c@{\hspace{2.0\tabcolsep}}c@{\hspace{2.0\tabcolsep}}c@{\hspace{2.0\tabcolsep}}c}
		\hline
		{} & $\Theta_{0}$ & $\Phi_{0}$ & $B_{0}$ & $H$ & $p_{0}$ & $\tau_\mathrm{exp}$ & $\chi^{2}_\mathrm{dir}$ & $\chi^{2}_\mathrm{mag}$ \\
		\hline
		Bepi & $39^{\circ}$ & $46^{\circ}$ & 82~nT & +1 & 0.23 & 20~h & 0.11 & 0.94 \\
        SolO & $68^{\circ}$ & $49^{\circ}$ & 38~nT & +1 & 0.06 & 74~h & 0.06 & 0.43 \\
        PSP  & $69^{\circ}$ & $167^{\circ}$ & 23~nT & +1 & 0.68 & 89~h & 0.15 & 0.06 \\
        ST-A & $14^{\circ}$ & $11^{\circ}$ & 15~nT & +1 & $-$0.28 & 57~h & 0.19 & 0.31 \\
		\hline
	\end{tabular}
\end{table}
%% -------------------------------------- %%

\subsection{Observations at BepiColombo} \label{subsec:bepi}

In-situ measurements at Bepi, located at ${\sim}0.44$~au during the event, are displayed in Figure~\ref{fig:obs_insitu}(a). Magnetic field data are supplied by the Mercury Planetary Orbiter Magnetometer \citep[MPO-MAG;][]{heyner2021}, whilst no plasma moments are available during the period under investigation---and more generally during most of the cruise phase. The first CME signatures appear at Bepi by means of an interplanetary shock passage on 25 September at 01:46~UT (vertical line in Figure~\ref{fig:obs_insitu}(a)), characterised by a moderate magnetic field jump and a quasi-parallel nature (see Table~\ref{tab:shocks})---we remark that, due to the lack of plasma data, it is not possible to determine with certainty whether this feature is a full-fledged shock. The following sheath region displays elevated magnetic field magnitudes for approximately two-thirds of its duration and subsequently dips to lower-than-ambient values. The identified CME ejecta period (\editone{dotted and solid} grey shading in Figure~\ref{fig:obs_insitu}(a)) features a complex magnetic field profile, characterised by two separate peaks---possibly the signature of \editone{an encounter that cuts first through the flank of the frontal CME body and then through the leg} (as expected from the WSA--Enlil results shown in Appendix~\ref{app:enlil}) or the outcome of interaction of the 23 September 2021 CME with the preceding streamer blowout to its east (described in Section~\ref{subsec:corona}). In fact, Bepi is the easternmost observer with respect to the CME nose amongst the four spacecraft at the time of impact, hence both the encounter with a flank and/or leg, as well as the detection of interaction signatures between the two slow streamer-blowout CMEs, can be reasonably expected.

Within the overall CME ejecta interval (bounded by the dotted areas in Figure~\ref{fig:obs_insitu}(a)) we identify a region of smoother and rotating magnetic field vectors (\editone{especially in the north--south component;} solid area in Figure~\ref{fig:obs_insitu}(a)), which we attribute to the core flux rope. Fitting with the EFF model yields a right-handed, moderately inclined rope with a rather low impact parameter (see Table~\ref{tab:frfits}). We remark, however, that in this case the speed of the flux rope, being unavailable at Bepi, is not used as a fit-constraining input, but is rather an output of the fitting procedure---given the rather high (${\sim}500$~km$\cdot$s$^{-1}$) resulting leading edge speed compared to the CME speed in the corona (${\sim}390$~km$\cdot$s$^{-1}$), it is likely that a fit employing actual plasma data would have generated somewhat different results.

\subsection{Observations at Solar Orbiter} \label{subsec:solo}

In-situ measurements at SolO, located at ${\sim}0.61$~au during the event, are displayed in Figure~\ref{fig:obs_insitu}(b). Data are provided by the Magnetometer \citep[MAG;][]{horbury2020} for magnetic field and the Proton and Alpha particle Sensor (PAS) of the Solar Wind Analyser \citep[SWA;][]{owen2020} for plasma parameters. The interplanetary shock driven by the 23 September 2021 CME is observed at SolO on 25 September at 18:25~UT (vertical line in Figure~\ref{fig:obs_insitu}(b)) and is characterised by a quasi-parallel nature and moderate strength (both Mach numbers are below 2, see Table~\ref{tab:shocks}). After the passage of the following sheath region, which displays progressively increasing magnetic field magnitudes, we identify a CME ejecta interval with clear flux rope signatures (solid shading in Figure~\ref{fig:obs_insitu}(b)). Unfortunately, a concurrent data gap in the magnetic field and plasma measurements prevents us from determining the trailing edge of the structure, but a likely upper limit is indicated by the hatched area in in Figure~\ref{fig:obs_insitu}(b))---i.e., we do not expect the CME ejecta to extend past that point, displaying a flat speed profile in contrast to the decreasing trend visible within the flux rope interval.

The magnetic field profile within the ejecta features a ``classic'' asymmetry in magnitude skewed towards the front of the rope, indicating CME expansion during propagation \citep[e.g.,][]{demoulin2009, nieves-chinchilla2018a}. In this case, the flux rope fitting is performed throughout the CME ejecta, from its leading edge up to the data gap, since flux rope signatures are displayed over the whole interval. The EFF model yields a central encounter with a right-handed, highly-inclined rope (see Table~\ref{tab:frfits}). Despite the fitting results appearing visually ``good'' (Figure~\ref{fig:obs_insitu}(b)), we remark that the very trailing portion of the ejecta (of unknown duration) is not included in the calculation, hence the real flux rope axial inclination may have been somewhat different than in our results.

\subsection{Observations at Parker Solar Probe} \label{subsec:psp}

In-situ measurements at PSP, located at ${\sim}0.78$~au during the event, are displayed in Figure~\ref{fig:obs_insitu}(c). Data are collected by the fluxgate magnetometer part of the FIELDS \citep{bale2016} instrument and the Solar Probe Cup \citep[SPC;][]{case2020} part of the Solar Wind Electrons Alphas and Protons \citep[SWEAP;][]{kasper2016} investigation. The sequence of events begins with a clear interplanetary shock passing the spacecraft on 26 September 2021 at 08:50~UT (vertical line in Figure~\ref{fig:obs_insitu}(c)), characterised by a quasi-perpendicular nature and higher strength than at SolO in terms of speed and Mach numbers (see Table~\ref{tab:shocks}). The magnetic field magnitude in the sheath region displays the most symmetric profile amongst the four spacecraft. Ejecta signatures (\editone{dotted and solid} shaded area in Figure~\ref{fig:obs_insitu}(c)) follow immediately after; however, besides the central portion that displays clear flux rope characteristics and a symmetric magnetic field magnitude profile (solid shading in Figure~\ref{fig:obs_insitu}(c)), the outer regions appear to preserve the rotation in the north--south direction but with highly-fluctuating fields, possibly due to erosion of the original rope \editone{due to reconnection with the ambient solar wind} \citep[e.g.,][]{lavraud2011, ruffenach2012}.

Flux rope fitting of the core flux rope interval yields a right-handed, highly-inclined rope that is crossed significantly far from its central axis (see Table~\ref{tab:frfits}). We also note that the resulting rope axis is oriented rather close to the Sun--spacecraft line, the separation between the two being only $13^{\circ}$ in longitude. We remark that possible erosion of the outer layers of the original ejecta, due to interactions with e.g.\ the ambient solar wind, may have altered the overall flux rope structure and orientation during transit. Finally, we note a region of high-density solar wind (but rather low field magnitudes) right after the passage of the ejecta trailing edge, possibly representing the accumulation of post-CME flows \citep[e.g.,][]{webb2016} given that there is no indication of a sector boundary crossing until ${\sim}12$~h later (note the longitudinal direction of the magnetic field being preserved before and after the CME passage).

\subsection{Observations at STEREO-A} \label{subsec:stereo}

In-situ measurements at STEREO-A, located at ${\sim}0.96$~au during the event, are displayed in Figure~\ref{fig:obs_insitu}(d). Time series come from the Magnetic Field Experiment \citep[MFE;][]{acuna2008}, part of the In situ Measurements of Particles And CME Transients \citep[IMPACT;][]{luhmann2008} suite, as well as the Plasma and Suprathermal Ion Composition \citep[PLASTIC;][]{galvin2008} investigation. We identify the interplanetary shock passage on 27 September 2021 at 01:51~UT (vertical line in Figure~\ref{fig:obs_insitu}(d)), and characterise it of quasi-parallel nature and associated with a moderate-to-high magnetic field jump (see Table~\ref{tab:shocks}). The sheath region, similarly to SolO measurements, displays and increasing profile in magnetic field magnitude. We find flux rope signatures throughout the CME ejecta interval (shaded area in Figure~\ref{fig:obs_insitu}(b)), but note that the magnetic field magnitude is characterised by a double-peak profile and that the interface between the two peaks displays sharp discontinuities in all magnetic field components.

Flux rope fitting with the EFF model yields a right-handed, lowly-inclined structure with its axis separated only by $11^{\circ}$ in longitude from the Sun--spacecraft line and that is encountered at intermediate distances from its central axis (see Table~\ref{tab:frfits}). We remark that this fit is associated with the largest error in the magnetic field components, likely due to the discontinuities mentioned above. Again, we note a region of high-density wind immediately following the CME ejecta, which may correspond to the similar structure found in PSP data.

%% -------------------------------------- %%
%  HINDCASTING WITH OSPREI
%% -------------------------------------- %%

\section{CME hindcasting with OSPREI} \label{sec:osprei}

Here, we present our efforts to simulate the 23 September 2021 CME in a hindcast fashion with the OSPREI\footnote{\url{https://github.com/ckay314/OSPREI}} modelling suite, which consists of three coupled modules: the Forecasting a CME's Altered Trajectory \citep[ForeCAT;][]{kay2015} that models CME deflections and rotations in the corona, the Another Type of Ensemble Arrival Time Results \citep[ANTEATR;][]{kay2022b} that propagates the CME through interplanetary space and includes the formation of a CME-driven sheath, and the ForeCAT In situ Data Observer \citep[FIDO;][]{kay2017} that generates synthetic in-situ profiles (along a time-dependent spacecraft trajectory or any point of choice). For more information on each module and additional technical details regarding OSPREI, we refer the reader to \citet{kay2022a}. OSPREI is a computationally-efficient analytical CME propagation model, which means that, despite the simplified physics compared to e.g.\ MHD calculations, it can be run with rapid turnaround in ensemble mode \citep[e.g.,][]{kay2018}. Hence, in this work, we first design the so-called `seed' run and evaluate its predicted impacts at the four probes, and then take advantage of a large ensemble (with 200 members) around this baseline run to evaluate the input parameters' influence over the variation in the field and plasma profiles generated at each spacecraft, both individually and together.

\subsection{Designing the `seed' run} \label{subsec:seed}

\begin{figure}
\centering
\includegraphics[width=.99\linewidth]{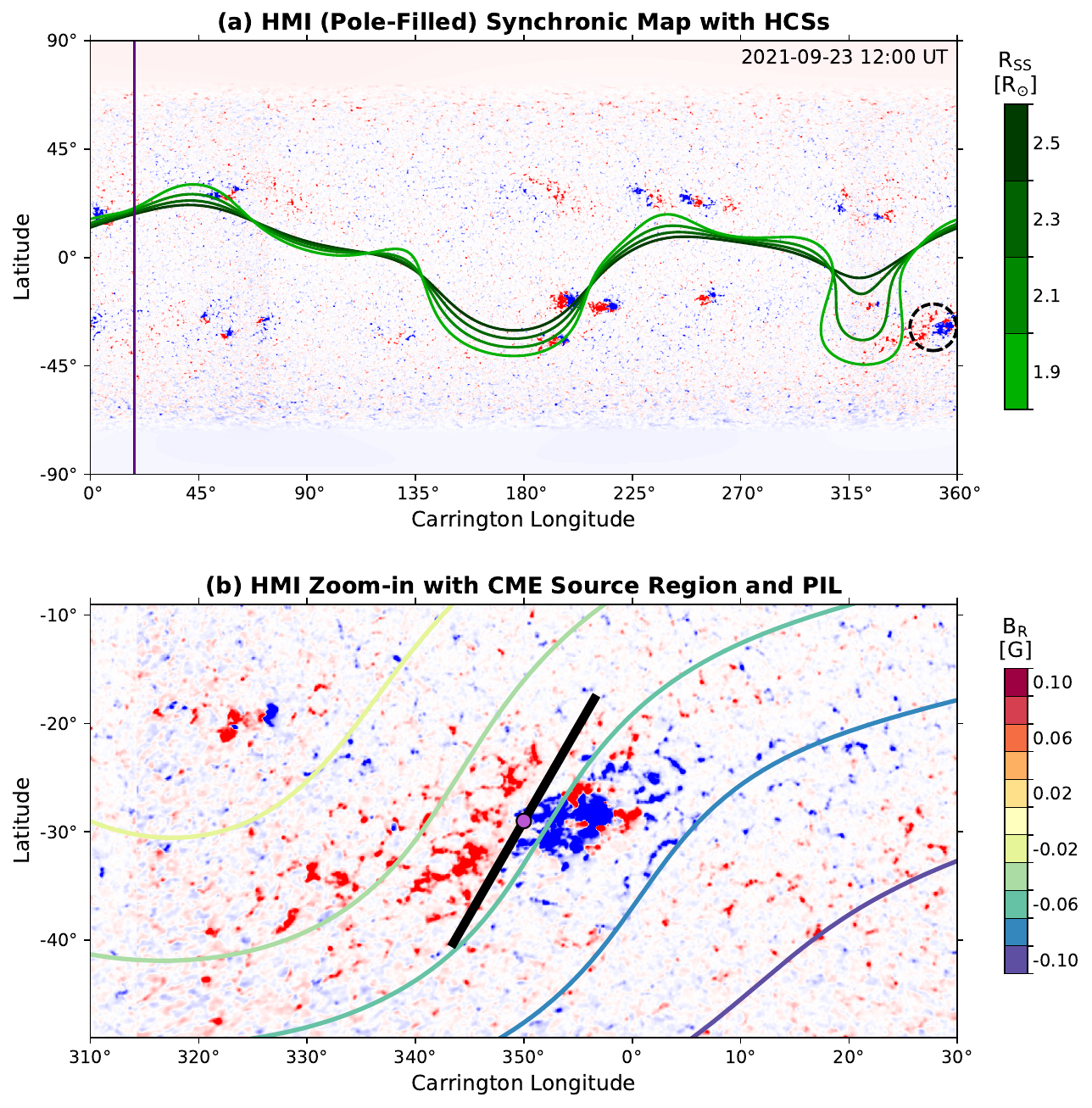}
\caption{Input photospheric conditions employed for the OSPREI simulation. (a) HMI (pole-filled) synchronic map for 2021-09-23 12:00~UT with the HCS resulting from four different PFSS source surface heights overlaid \editone{in shades of green}. The magnetogram has been saturated to ${\pm}100$~G, with positive (negative) field shown in red (blue). The source region of the 23 September 2021 event is circled in black, and the Carrington longitude of Earth at the time of the eruption (04:30~UT) is marked with a purple vertical line. (b) Zoom-in on the source region of the 23 September 2021 CME, showing PFSS contours ($R_\mathrm{SS} = 2.5\,R_{\odot}$) of the radial magnetic field as well as the location of the flux rope nose (purple dot) with its associated PIL (black line).}
\label{fig:magnetogram}
\end{figure}

The first step towards modelling the coronal and heliospheric propagation of CMEs with the OSPREI suite is to define the photospheric boundary conditions. Since the CME of interest erupted ${\sim}30^{\circ}$ away from the Sun--Earth line in longitude, we employ in this study the pole-filled SDO/HMI synchronic map for 23 September 2021, which is generated by replacing from a standard Carrington map daily observations within ${\pm}60^{\circ}$ of the central meridian as seen from Earth \citep{hayashi2015}. Using this input magnetogram, the coronal conditions---i.e., the coronal magnetic field---are generated by applying the Potential Field Source Surface \citep[PFSS;][]{wang1992} model. Since it has been shown that the choice of PFSS source-surface radius ($R_\mathrm{SS}$) can have more or less prominent effects on the CME evolution modelled by OSPREI \citep[see][]{ledvina2023}, we select four source surfaces to test for significant differences in the seed run---the ``classic'' $2.5\,R_{\odot}$ \citep{altschuler1969} as well as the lower heights of 2.3, 2.1, and $1.9\,R_{\odot}$. The input magnetogram, the location of the heliospheric current sheet (HCS) resulting from the different PFSS source surfaces, as well as the source region of the 23 September 2021 CME are shown in Figure~\ref{fig:magnetogram}(a).

%% -------------------------------------- %%
%% Figure: The magnetogram
%% -------------------------------------- %%

%% -------------------------------------- %%
%% Table: OSPREI inputs
\begin{table}
	\centering
	\caption{OSPREI input parameters for the seed run and variations employed for the ensemble run. The parameters displayed are, from top to bottom: eruption date and time, radial distance at which coronal propagation (ForeCAT) ends ($R_\mathrm{FC}$), flux rope initial height ($R_{0}$), latitude ($\theta_{0}$), Carrington longitude ($\phi_{0}$), and tilt ($\psi_{0}$), helicity sign ($H$), axial magnetic field strength ($B_\mathrm{FR}$), CME mass ($M_\mathrm{FR}$), CME temperature ($T_\mathrm{FR}$), face-on (AW) and edge-on (AW$_{\perp}$) half-angular width, axial ($\delta_\mathrm{AX}$) and cross-sectional ($\delta_\mathrm{CS}$) aspect ratio, initial slow-rise speed ($V_{0}$), altitude at which kinematics transition from slow rise to rapid acceleration ($a_\mathrm{0}$), maximum coronal speed ($V_{1}$), altitude at which kinematics transition from rapid acceleration to constant speed ($a_\mathrm{1}$), flux rope adiabatic index ($\gamma$), interplanetary expansion factor ($f_\mathrm{exp}$), heliocentric distance used for the background wind description ($R_\mathrm{SW}$), ambient drag coefficient ($C_{d}$), as well as solar wind speed ($V_\mathrm{SW}$), magnetic field magnitude ($B_\mathrm{SW}$), density ($N_\mathrm{SW}$), and temperature ($T_\mathrm{SW}$).}
	\label{tab:osprei}
	\begin{tabular}{l@{\hspace{8.0\tabcolsep}}l@{\hspace{8.0\tabcolsep}}l}
		\hline
		{} & Seed & Ensemble \\
		\hline
		Date & 2021-09-23 & --- \\
        Time & 04:30 UT & --- \\
        $R_\mathrm{FC}$ & 20\,$R_{\odot}$ & --- \\
        $R_{0}$ & 1.2\,$R_{\odot}$ & --- \\
        $\theta_{0}$ & $-29^{\circ}$ & ${\pm}3^{\circ}$ \\
        $\phi_{0}$ & $350^{\circ}$ & ${\pm}5^{\circ}$ \\
        $\psi_{0}$ & $60^{\circ}$ & ${\pm}10^{\circ}$ \\
        $H$ & +1 & --- \\
        $B_\mathrm{FR}$ & $2.0 \times 10^{3}$ nT & ${\pm}0.5 \times 10^{3}$ nT\\ 
        $M_\mathrm{FR}$ & $1.0 \times 10^{16}$ g & ${\pm}0.5 \times 10^{16}$ g\\
        $T_\mathrm{FR}$ & $1.5 \times 10^{5}$ K & ${\pm}0.5 \times 10^{5}$ K \\ 
        AW & $36^{\circ}$ & ${\pm}5^{\circ}$ \\
        AW$_{\perp}$ & $15^{\circ}$ & ${\pm}2^{\circ}$ \\
        $\delta_\mathrm{AX}$ & 0.7 & ${\pm}0.1$ \\
        $\delta_\mathrm{CS}$ & 0.9 & ${\pm}0.1$ \\
        $V_{0}$ & 50~km$\cdot$s$^{-1}$ & $\pm 20$~km$\cdot$s$^{-1}$ \\
        $a_\mathrm{0}$ & 1.7\,$R_{\odot}$ & ${\pm}0.1\,R_{\odot}$ \\
        $V_{1}$ & 390~km$\cdot$s$^{-1}$ & $\pm 50$~km$\cdot$s$^{-1}$ \\
        $a_\mathrm{1}$ & 8.0\,$R_{\odot}$ & ${\pm}1.0\,R_{\odot}$ \\
        $\gamma$ & 1.33 & ${\pm}0.1$ \\
        $f_\mathrm{exp}$ & 0.5 & ${\pm}0.1$ \\
        $R_\mathrm{SW}$ & 213\,$R_{\odot}$ & --- \\
        $C_{d}$ & 1.0 & ${\pm}0.25$ \\ 
        $V_\mathrm{SW}$ & 340~km$\cdot$s$^{-1}$ & ${\pm}40$~km$\cdot$s$^{-1}$\\ 
        $B_\mathrm{SW}$ & 5 nT & ${\pm}3$ nT\\ 
        $N_\mathrm{SW}$ & 10 cm$^{-3}$ & ${\pm}5$ cm$^{-3}$\\ 
        $T_\mathrm{SW}$ & $6.0 \times 10^{4}$ K & ${\pm}1.0 \times 10^{4}$ K\\
		\hline
	\end{tabular}
\end{table}
%% -------------------------------------- %%

%% -------------------------------------- %%
%% Figure: Evolution of the seed run
\begin{figure*}
\centering
\includegraphics[width=.97\linewidth]{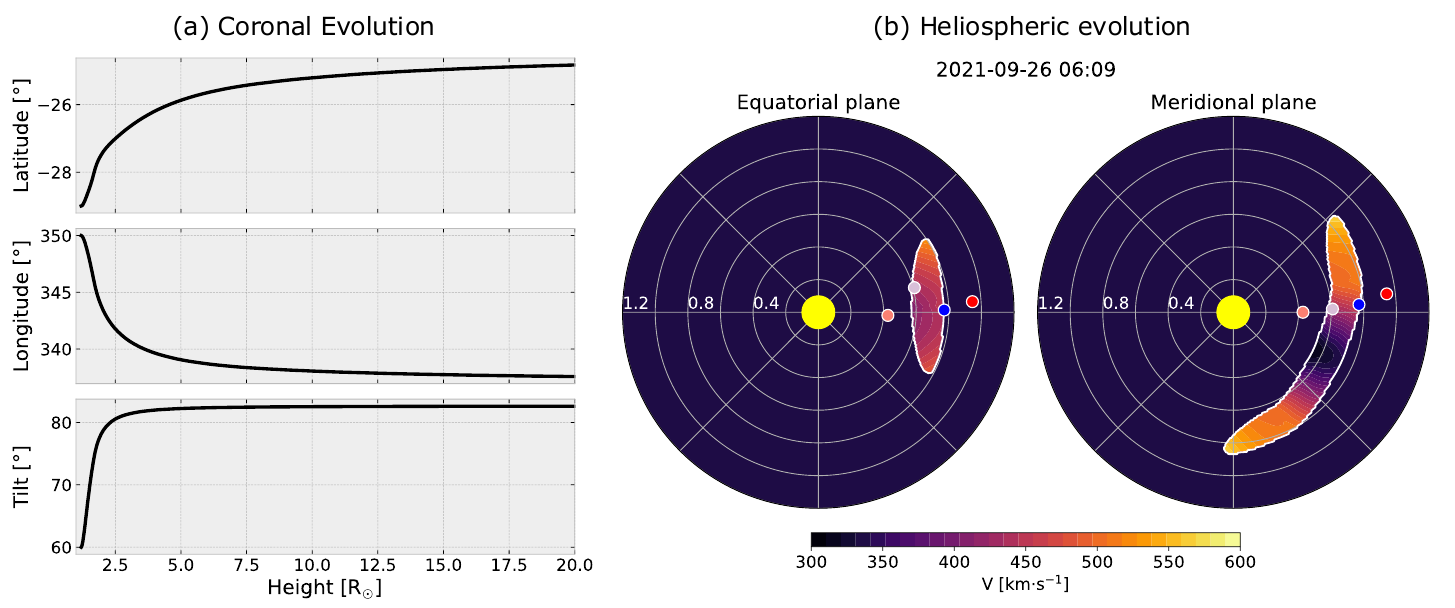}
\caption{Overview of the (a) coronal and (b) heliospheric evolution of the CME modelled as the seed run for OSPREI. (a) ForeCAT deflections and rotations up to 20\,$R_{\circ}$. (b) Snapshot of the CME evolution in interplanetary space as seen from (left) the equatorial and (right) the nose-centred meridional planes. The quantity shown is the solar wind bulk speed, and the four circles represent the positions of the four spacecraft (by increasing distance from the Sun, Bepi, SolO, PSP, and STEREO-A) projected onto the respective planes.}
\label{fig:seed_evo}
\end{figure*}
%% -------------------------------------- %%

Once the background conditions have been generated, we begin defining the flux rope input parameters, which are listed in detail in Table~\ref{tab:osprei}. In the current implementation, OSPREI employs the Elliptic-Cylindrical \citep[EC;][]{nieves-chinchilla2018b} flux rope model to describe the CME morphology and magnetic configuration. We select the CME initial latitude ($\theta_{0}$), longitude ($\phi_{0}$), and tilt ($\psi_{0}$) based on observations of the active region and its associated PIL (see Figure~\ref{fig:magnetogram}(b)). The helicity sign (or chirality, $H$) is assumed positive based on remote-sensing observations of the eruption \citep[see][]{palmerio2017}, in particular from the right-skewness of the PEAs with respect to the underlying PIL. Values for CME internal magnetic field ($B_\mathrm{FR}$), mass ($M_\mathrm{FR}$), and temperature ($T_\mathrm{FR}$) are set based on best-guess approximations. Parameters describing CME morphology (AW, AW$_{\perp}$, $\delta_\mathrm{AX}$, and $\delta_\mathrm{CS}$) are loosely based on the GCS reconstructions shown in Figure~\ref{fig:obs_wl} and include modifications necessary to grant stability of the EC solution---for example, by elongating the CME half-width along its central axis, yielding a more cylindrical structure against the ``rounded'' ellipsoid employed for the WSA--Enlil run (see Appendix~\ref{app:enlil}). Properties characterising CME kinematics in the corona ($V_{0}$, $a_\mathrm{0}$, $V_{1}$, and $a_\mathrm{1}$) are selected based on off-limb WL observations from SOHO/LASCO. Parameters governing CME propagation in interplanetary space ($\gamma$ and $f_\mathrm{exp}$) are left to their default values. Finally, the background solar wind conditions are defined at the STEREO-A position (213\,$R_{\odot}$) and scaled with heliocentric distance accordingly to the other locations. The drag coefficient \citep[$C_{d}$; e.g.,][]{vrsnak2013} is kept at its default value of one. The values for magnetic field ($B_\mathrm{SW}$), density ($N_\mathrm{SW}$), and temperature ($T_\mathrm{SW}$) are taken directly from in-situ measurements preceding the CME arrival, whilst the solar wind bulk velocity ($V_\mathrm{SW}$) is slightly lowered from its STEREO-A values to compensate from the significantly lower speeds found at SolO (note that we are considering a uniform, constant solar wind background speed in this work).

We test the seed run input parameters reported in Table~\ref{tab:osprei} using the four $R_\mathrm{SS}$ values shown in Figure~\ref{fig:magnetogram}(a) for the PFSS solution, and find no significant differences in the results. Hence, this particular configuration does not appear to be affected by the PFSS source-surface radius in a notable way, and in the following we shall consider only the ``classic'' $R_\mathrm{SS} = 2.5\,R_{\odot}$. And overview of the coronal and heliospheric evolution of the 23 September 2021 CME modelled in the seed run is presented in Figure~\ref{fig:seed_evo} and in Supplementary Video~3. In the coronal domain (i.e., the region ${\leq}20\,R_{\odot}$ where ForeCAT operates), the CME is seen to deflect towards the northeast and to rotate to higher inclinations, its axial parameters changing from ($\theta_{0}$, $\phi_{0}$, $\psi_{0}$) = ($-29^{\circ}$, $350^{\circ}$, $60^{\circ}$) to ($\theta$, $\phi$, $\psi$) = ($-24.8^{\circ}$, $337.6^{\circ}$, $82.6^{\circ}$) as shown in Figure~\ref{fig:seed_evo}(a). In fact, in interplanetary space the CME appears significantly more extended in latitude than in longitude, as can be seen in Figure~\ref{fig:seed_evo}(b) and Supplementary Video~3. The four spacecraft of interest (Bepi, SolO, PSP, and STEREO-A) are all impacted by the CME north of its nose, with a closer-to-flank encounter at Bepi and a more central one at the remaining three probes.

\begin{figure*}
\centering
\includegraphics[width=.95\linewidth]{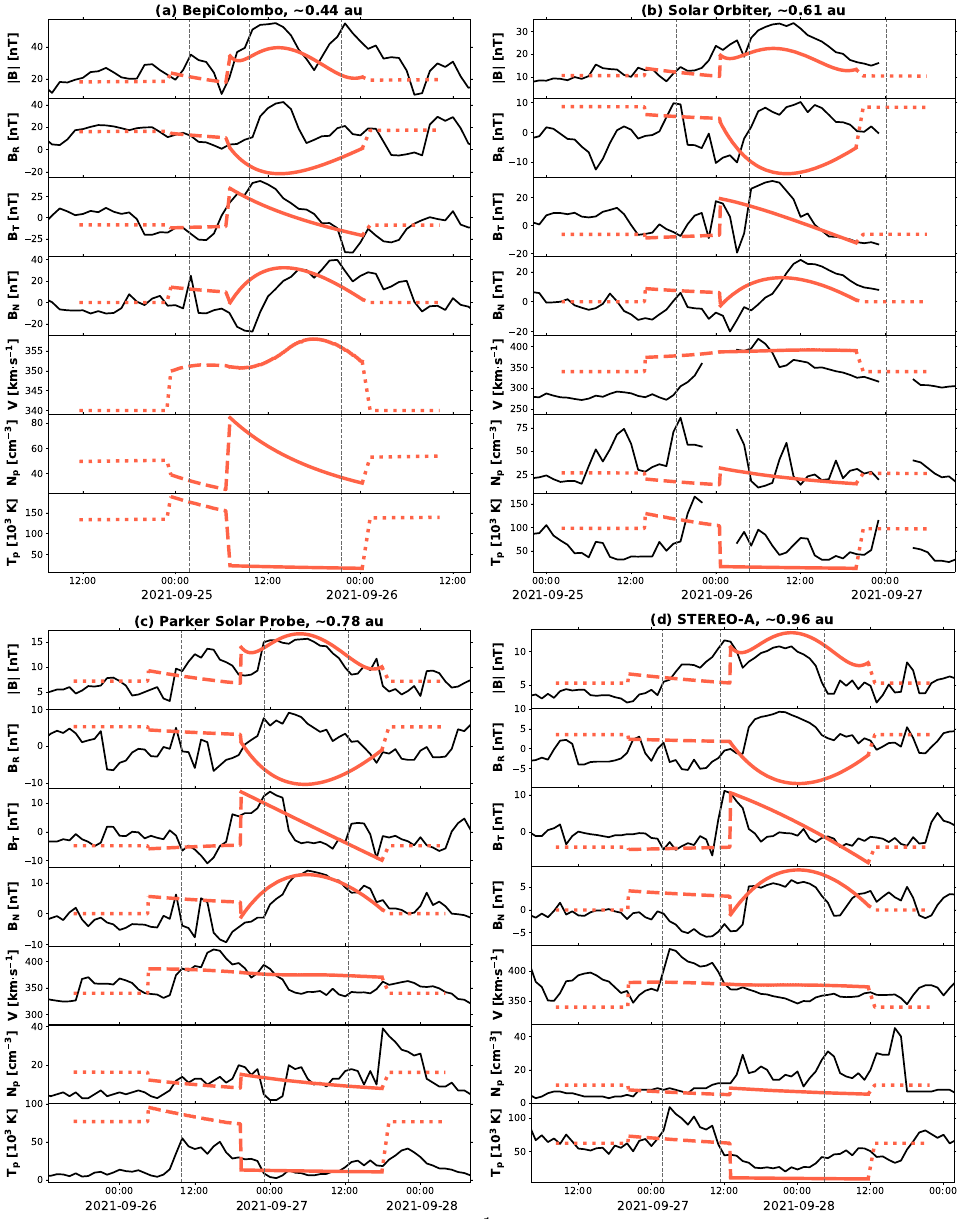}
\caption{Overview of the OSPREI seed simulation run results shown against in-situ measurements of the 23 September 2021 CME at (a) Bepi, (b) SolO, (c) PSP, and (d) STEREO-A. Each plot shows, from top to bottom: magnetic field magnitude, magnetic field Cartesian components in RTN coordinates, solar wind bulk speed, as well as proton density and temperature. The spacecraft data are displayed in black, whilst the OSPREI synthetic profiles are shown in orange for (dotted) background wind, (dashed) sheath region, and (solid) CME ejecta intervals. The \editone{dashed} vertical lines mark \editone{(leftmost)} the shock arrival and \editone{(remaining two)} the flux rope boundaries as observed by each spacecraft.}
\label{fig:seed_insitu}
\end{figure*}

The OSPREI synthetic in-situ profiles compared to spacecraft measurements at each location are shown in \editone{Figure~\ref{fig:seed_insitu}}. The modelled arrival times of shocks/sheaths and ejecta leading edges are all within a few hours of the observed ones, i.e.\ within the well-known uncertainties (${\sim}$10~h) associated with predictions of CME propagation \citep{kay2024b}. The magnetic field magnitude is underestimated at Bepi and SolO, in agreement with in-situ measurements at PSP, and slightly overestimated at STEREO-A. The individual field components appear to follow the overall west-to-east rotation in $B_{T}$ and the largely-positive nature of $B_{N}$, but display opposite sign in $B_{R}$ compared to observations. In the next section, this seed run is used as the basis for an ensemble run that evaluates the ``best fit'' at each spacecraft both separately and when considered together.

%% -------------------------------------- %%
%% Figure: The seed run in situ
%% -------------------------------------- %%

\subsection{Ensemble modelling} \label{subsec:ensemble}

A number of studies have shown that predictions of CME magnetic configuration and/or arrival time can depend more or less strongly on the choice of CME input parameters, even within a single model \citep[e.g.,][]{kay2021, palmerio2022a}. To evaluate variations in the synthetic in-situ profiles due to inputs and to determine differences between individual runs producing the best matches at each spacecraft separately and together, we now consider fluctuations around the seed simulation and run OSPREI in its ensemble mode. We employ a relatively large ($N_\mathrm{ens}$ = 200) number of ensemble members, and the ranges of the variations applied on the input parameters of the seed run are reported in Table~\ref{tab:osprei}. The $N_\mathrm{ens}$ = 200 choice is somewhat arbitrary, but reasonably samples the assumed intervals of all the varied parameters. \editone{The allowed ranges of variations for each parameter are set to be more conservative (i.e., larger) than what is usually assumed in OSPREI \citep[e.g.,][]{palmerio2021b, kay2022a}, to account for the inherent complexity in characterising both the source region and coronal evolution for the CME under study (see Section~\ref{sec:remote}).} In the post-processing phase of the ensemble simulation, we use spacecraft observations at the four locations to compute metrics and thus obtain a ``best fit'' for each probe considered both individually and collectively. The goodness-of-fit score is defined as the sum of the fractional mean absolute error of the hourly averages for each magnetic field and plasma parameter together with a timing error consisting of the absolute error in days for all three CME bounds, i.e.\ shock, ejecta leading edge, and ejecta trailing edge arrivals \citep[see also][]{kay2017}. Within each ensemble member, we also sum the goodness-of-fit scores across each spacecraft to compute a ``global'' metric that aims to identify the run that best reproduces the CME behaviour at the four locations when considered together.

%% -------------------------------------- %%
%% Figure: Evolution of the ensemble run
\begin{figure*}
\centering
\includegraphics[width=.97\linewidth]{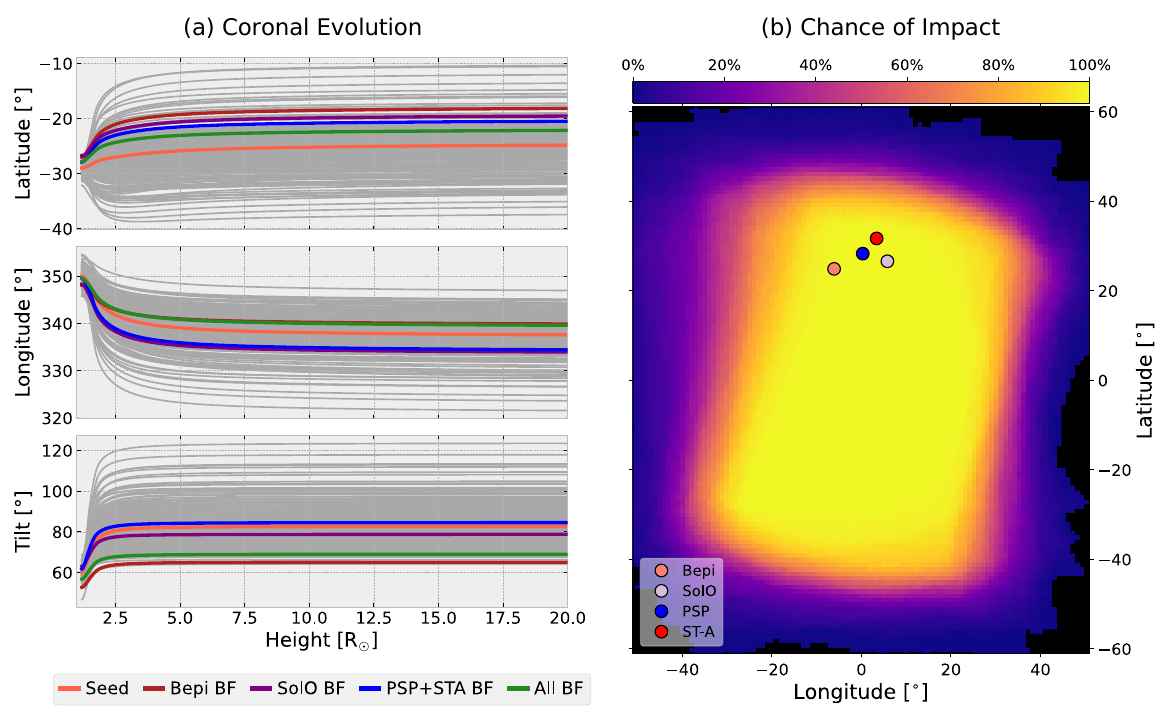}
\caption{Overview of the (a) coronal and (b) heliospheric evolution of the (200-member) ensemble CME run modelled with OSPREI. (a) ForeCAT deflections and rotations up to 20\,$R_{\circ}$. The seed run as well as the various best-fit runs are highlighted in different colours, whilst the remaining ensemble members are shown in grey. (b) Radial slice of the CME angular extent at the time of impact at PSP, shown as a heat map representing the chance of impact at a given latitude--longitude coordinate across the ensemble. The superposed circles mark the projected positions of the four spacecraft at their respective observed CME arrival times (i.e., the times shown in Table~\ref{tab:positions}). The figure is displayed in the frame centred at the CME nose of the seed run.}
\label{fig:ens_evo}
\end{figure*}
%% -------------------------------------- %%

An overview of the coronal and heliospheric evolution of the ensemble run is provided in Figure~\ref{fig:ens_evo}. The coronal evolution of the different ensemble members (Figure~\ref{fig:ens_evo}(a)) shows an approximately symmetric (${\pm}15^{\circ}$) dispersion in latitude around the seed run, whilst the ensemble distribution in longitude and tilt have developed a distinct asymmetry by $20\,R_\odot$. The longitude has an extended tail below the seed value and is condensed above it ([$-20^{\circ}, +10^{\circ}$]), whereas the tilt angle shows the opposite trend ([$-20^{\circ}, +40^{\circ}$]). The symmetry found in the latitudinal evolution suggests that the seed is in a relatively balanced location with respect to the magnetic forces. We would expect the global forces to push the CME north towards the HCS, and the local forces to depend on the exact location within the AR. Small changes in the initial position can affect the balance between these forces either way. The persistent eastward motion is likely in large part due to the strong negative polarity region of the AR, but we would also expect some eastward motion from the global forces since that is the direction of the closest part of the HCS. In terms of evolution of the best-fit runs with respect to the seed, we note that all of them are characterised by higher latitudes throughout the ForeCAT domain, whilst mixed patterns are visible in the remaining two parameters. In the longitude, the global best-fit run is slightly westwards of the seed, but the single-spacecraft best-fit runs are all clustered a few degrees eastwards of the seed---note that a single run gives the best fit at both PSP and STEREO-A, hence they are considered together throughout this analysis. In the axial tilt, the single-spacecraft best-fit ensemble members for SolO and PSP+STEREO-A are clustered close to the seed, whilst the Bepi and global best-fit run are less tilted than the seed by ${\sim}15^{\circ}$. The distribution in coronal evolution evident from Figure~\ref{fig:ens_evo}(a) results in the CME affecting slightly different angular wedges during its heliospheric propagation across the ensemble, as shown in the heat map of Figure~\ref{fig:ens_evo}(b). Note that, in OSPREI, the CME size increases with distance as a result of magnetic and thermal forces between the ejecta and the solar wind background, hence the radial slice shown in the figure is arbitrarily chosen at the corresponding impact time at PSP for each ensemble member. The projected positions (at each corresponding arrival time) of the four spacecraft over the CME angular extent show that all probes are located in a region of high chance of impact (>90\%) and that all encounters take place north of the CME nose, whilst the wider distribution in longitude indicates that it depends on the specific run whether a probe crosses the CME east or west of its highly-inclined symmetry axis.

%% -------------------------------------- %%
%% Figure: Ensemble run in situ
\begin{figure*}
\centering
\includegraphics[width=.945\linewidth]{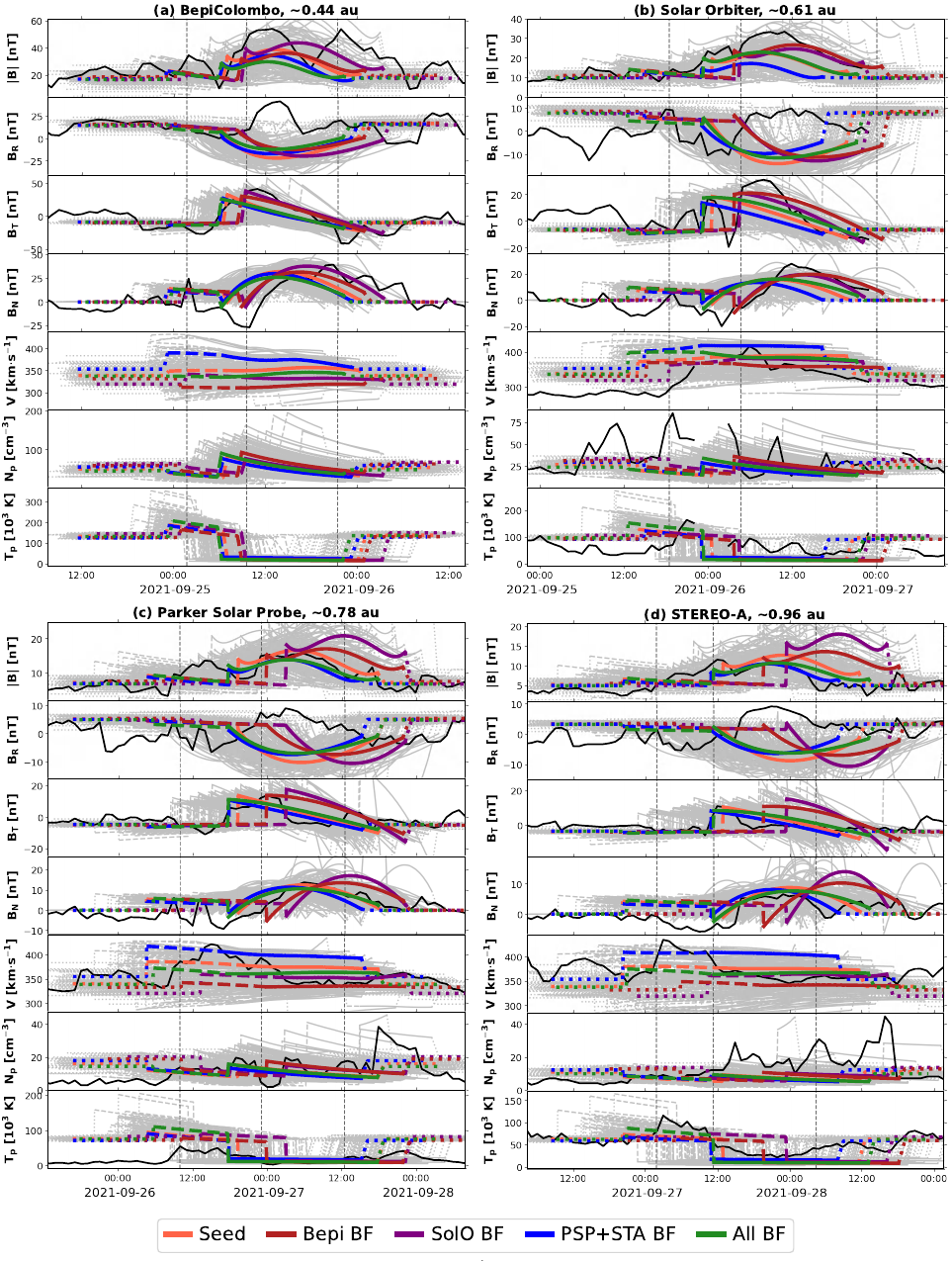}
\caption{Overview of the OSPREI (200-member) ensemble simulation run results shown against in-situ measurements of the 23 September 2021 CME at (a) Bepi, (b) SolO, (c) PSP, and (d) STEREO-A. The seed run as well as the various best-fit runs are highlighted in different colours, whilst the remaining ensemble members are shown in grey. All panels and parameters are shown in the same format as Figure~\ref{fig:seed_insitu}.}
\label{fig:ens_insitu}
\end{figure*}
%% -------------------------------------- %%

The synthetic in-situ profiles resulting from the ensemble run are shown in Figure~\ref{fig:ens_insitu}, \editone{in the same colour-coding as Figure~\ref{fig:ens_evo}}. First of all, we note that most shock/CME arrivals take place within a few hours of the seed but a few runs display larger differences (>12~h), indicating that different combinations of CME initial speed, CME acceleration profile, and/or drag coefficient may result in significantly different predictions even in the case of a fixed, uniform solar wind background. In terms of magnetic configuration, despite more or less prominent differences in the field magnitude all runs tend to follow the seed's overall trend of negative $B_{R}$, positive-to-negative rotating $B_{T}$, and mostly positive $B_{N}$. This is not surprising, since the CME tilt over the ensemble is spread over approximately ${\pm}30^{\circ}$ around the north direction ($90^{\circ}$) and all encounters take place north of the CME nose. With respect to the seed, the single-spacecraft best-fit runs tend to show a better alignment of the ejecta leading edge with observations with the exception of PSP, where the additional rotation in $B_{T}$ and $B_{N}$ preceding what we defined as the flux rope interval (see Section~\ref{subsec:psp} and Figure~\ref{fig:obs_insitu}(c)) appears to be affecting fitting results. The global best-fit profiles are remarkably similar to the single-spacecraft ones at PSP and STEREO-A, whilst larger differences are present at Bepi and SolO\editone{---this may be due to the two outer probes being characterised by the same single-spacecraft best fit, thus carrying a larger weight in the global best-fit calculation, but also being located between the two inner probes in longitude, thus ``averaging out'' differences due to angular separation}. Overall, the seed and best-fit runs do not show particularly stark differences at any of the locations considered, and all appear approximately in the middle of the ensemble distribution for each curve.

However, despite each best-fit run being comfortably within the ensemble distribution at each of the four spacecraft individually, the best-fit ensemble member for a particular inner spacecraft (Bepi and SolO) has no guarantee of it being the best or even a reasonable fit at any of the further spacecraft (PSP and STEREO-A). This is clearly illustrated by the maroon and purple curves shown in Figures~\ref{fig:ens_insitu}(c--d), which are the best-fit ensemble members for Bepi and SolO propagated to the PSP and STEREO-A positions, respectively. The set of best-fit profiles at Bepi and SolO show relatively minor timing, shape, and magnitude differences between them whereas by the radial distances of PSP and STEREO-A, both the Bepi and SolO ensemble members produce magnetic field profiles that are more prominent outliers in their respective distributions and represent the observational in-situ data less adequately. 

%% -------------------------------------- %%
%  DISCUSSION
%% -------------------------------------- %%

\section{Discussion} \label{sec:discussion}

Observations of the 23 September 2021 CME (Sections~\ref{sec:remote} and \ref{sec:insitu}) as well as our modelling efforts with OSPREI (Section~\ref{sec:osprei}) have shown that the event considered in this work presents many complex characteristics that make both interpretation of the spacecraft measurements and prediction of its heliospheric impact(s) particularly challenging. Here, we synthesise the results and findings presented in the previous sections to build a comprehensive overview of the CME's Sun-to-1~au transit and large-scale structure whilst still highlighting the complexities and disagreements where they arise.

\subsection{The observational perspective} \label{subsec:discobs}

The 23 September 2021 eruption is characterised by a complex, \editone{nested-AR source region \citep[][]{karpen2024}}, multi-stage eruption dynamics, and ambiguous WL signatures. \editone{The first (precursor) stage involves a circular ribbon flare in the closed flux region near the edge of the helmet streamer belt \citep{wyper2016b}. The second (main) stage of the eruption generates the CME with a classic two-ribbon flare, which is also associated with simultaneous twin coronal dimmings \citep{thompson1998}.} The CME-producing PEA and its underlying PIL's north--south orientation gives an estimated CME flux rope orientation of west--north--east \citep[WNE; following][]{bothmer1998, mulligan1998} for a right-handed flux rope (see also Figure~\ref{fig:magnetogram}(b)), which is consistent with the large-scale orientation of the overlying helmet streamer belt. It is not possible to observe in detail the CME deflecting and/or rotating in the corona \citep[e.g.,][]{vourlidas2011} due to its faint appearance and because of the presence of simultaneous eruptions in the available WL data. Nevertheless, one significant point of interest of this event is that it was encountered in situ by four probes approximately equally distributed in heliocentric distance between 0.4 and 1~au; hence, we shall leverage these measurements to compare the estimated CME structure at the Sun with that observed in situ, and at the same time to evaluate similarities and differences amongst the four in-situ data sets.

At each spacecraft, the first CME-related structure to be measured is the CME-driven shock. Despite the evident difficulties in addressing the shock properties due to the fact that no plasma data (more precisely, speed components) are available at Bepi and STEREO-A, we recover a compatible set of parameters at the different observers, with no strong variations in the compression ratios, shock speeds, and Mach numbers (see Table~\ref{tab:shocks}). The shock appears subcritical at both SolO and PSP, a property frequently observed in the case of interplanetary shocks \citep[e.g.][]{kilpua2015}. We note a level of variability in the local shock normals as shown in Table~\ref{tab:shocks}, indicating that local shock properties may vary more or less strongly at different heliospheric locations for a single event. In this regard, the exploitation of multiple heliospheric observers is an invaluable resource to quantify local variability against large-scale evolution effects \citep[see also][for recent efforts in defining CME-driven shock properties in multi-spacecraft data]{palmerio2024b, trotta2024a}.

The sheath region immediately following the shock displays different characteristics at the different probes, first and foremost in the profiles of the magnetic field magnitude---increasing towards the ejecta at SolO and STEREO-A, abruptly declining at Bepi, and more plateau-like at PSP (see Figure~\ref{fig:obs_insitu}). At least at the three spacecraft for which plasma data are available, the CME is embedded in the slow solar wind, hence these difference do not appear to be related to the eruption propagating through profoundly different local environments. Nevertheless, previous studies have shown that the structure of CME-driven sheaths may differ more or less significantly both at the local and global level even for multi-spacecraft encounters realised in radial alignment or in close proximity of the probes involved \citep[see, e.g.,][]{good2020, kilpua2021}. The measurements investigated here highlight further the importance of reaching a deeper understanding of the complex dynamics between the shock driver and the ambient medium responsible for sheath formation and evolution, especially given their potential to drive severe space weather effects \citep[e.g.,][]{pulkkinen2007}.

\begin{figure}
\centering
\includegraphics[width=.99\linewidth]{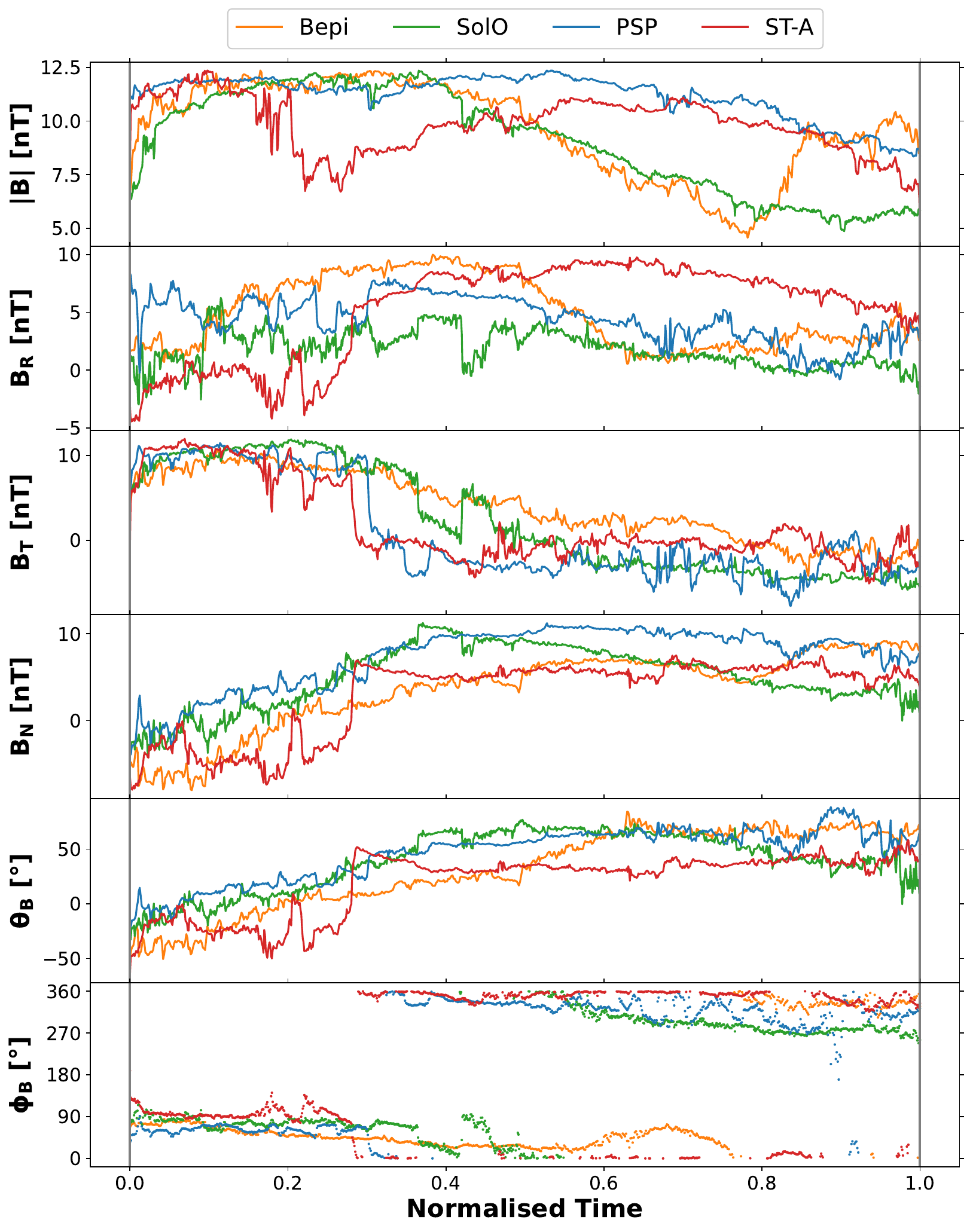}
\caption{Ejecta magnetic fields of the 23 September 2021 CME as observed by Bepi, SolO, PSP, and STEREO-A, \editone{normalised in duration so that each of the leading and trailing edges are aligned}. The time series for (from top to bottom) field strength, (RTN) Cartesian field components, and magnetic field angles \editone{in the latitudinal ($\theta_{B}$) and longitudinal ($\phi_{B}$) directions} have been scaled in magnitude as to superpose each spacecraft's measurements onto the STEREO-A ones. The scaling factor is obtained by normalising the maximum magnetic field magnitude at a given probe to the corresponding value at STEREO-A (note that this scaling does not apply to the \editone{field angles}). For each spacecraft data set, the interval shown corresponds to the flux rope interval (where flux rope fitting is performed, see Figure~\ref{fig:obs_insitu}).}
\label{fig:mag_stretch}
\end{figure}

Finally, all four probes are impacted by the CME ejecta, where we identify several similarities as well as differences across the different data sets. To aid in this analysis and discussion of its results, Figure~\ref{fig:mag_stretch} displays the four sets of ejecta magnetic field time series normalised in time, scaled in magnitude, and superposed onto the STEREO-A data \editone{(i.e., at the outermost spacecraft)}. It is clear that the $B_{T}$ and $B_{N}$ components follow a very similar trend across the four time series, whilst larger variability is present is $B_{R}$ and $|B|$. Visually, the positive-to-negative rotation in $B_{T}$ and the mostly-northwards $B_{N}$ (with some southwards fields at the ejecta front measured by Bepi and STEREO-A) suggest an overall flux rope type close to the WNE configuration estimated from remote-sensing observations. This is consistent with previous studies, which have shown that the majority of CMEs tend to maintain their axial orientation in interplanetary space within ${\pm}45^{\circ}$ of their solar counterpart \citep[e.g.,][]{palmerio2018, xie2021}. Nevertheless, this agreement at the global level is contrasted by some prominent local differences, the most striking of which being the profile of the magnetic field magnitude within the flux rope ejecta. The characteristic shape of the $|B|$ curves---having a distinct, often asymmetric peak offset towards the leading edge of the ejecta---suggests more central encounters at SolO and PSP and more complicated spacecraft crossings at Bepi and STEREO-A (cf.\ the more variable $|B|$ profiles), or alternatively a rope that is locally distorted. Given the overall agreement in the large-scale trend of the $B_T$ and $B_N$ components at all four spacecraft, the observed variability in $|B|$ appears to be largely due to the differences in each probe's $B_R$ profiles, e.g.\ large-scale $B_R$ polarity changes, HCS proximity, etc. This suggests that the $B_R$ component may contain important information about the relative position of the spacecraft with respect to the CME as well as the CME's orientation with respect to the interplanetary sector structure. Additionally, the flux rope fitting results (Table~\ref{tab:frfits}), despite all giving a westerly axial direction ($+\hat{T}$ component), suggest a high-inclination rope (of WNE type) at SolO and PSP, but a low-inclination one (of south--west--north or SWN type) at Bepi and STEREO-A. These fitting results reflect the difference(s) in each event's $B_{T}$ and $B_{N}$ profiles. For example, the low-inclination encounters (Bepi and STEREO-A) have larger southern fields (more negative $B_{N}$) at the beginning of their CME intervals as well as shallower negative $B_{T}$ regions in the trailing half of the ejecta compared to the higher-inclination encounters at SolO and PSP. 

%% -------------------------------------- %%
%% Figure: Mag field stretching
%% -------------------------------------- %%

\subsection{The modelling perspective} \label{subsec:discmod}

Analytical modelling of the 23 September 2021 CME provides the rare opportunity to validate and compare results at four well-separated locations in the sub-au heliosphere: Bepi (0.44~au), SolO (0.61~au), PSP (0.78~au), and STEREO-A (0.96~au). Forward-modelling of CME propagation and magnetic structure with analytical codes has thus far focussed largely on two-spacecraft encounters realised in near-radial alignment \citep[e.g.,][]{mostl2018, sarkar2024}, with results suggesting that data collected by an inner probe should be used to constrain models for more accurate predictions at 1~au \citep[e.g.,][]{kubicka2016, laker2024}. In this work, we employed the OSPREI modelling suite to evaluate the coronal and heliospheric evolution of the 23 September 2021 CME and to evaluate its impact across interplanetary space via an ensemble approach. After determining that the seed run largely captures the overall structure of the CME at the different probes (see Figure~\ref{fig:seed_insitu}), we computed goodness-of-fit metrics for each ensemble member to extrapolate single-spacecraft and global ``best runs''.

One interesting finding of this analysis is that the various best-fit solutions show little spread at Bepi, and then progressively diverge with heliocentric distance up to STEREO-A, where the most prominent differences are displayed (see Figure~\ref{fig:ens_insitu}). This is intuitively reasonable, since the effects of CME evolution on in-situ profiles are expected to become more evident with distance from the Sun. Nevertheless, we remark that the CME modelled here with OSPREI propagates through a constant, uniform solar wind background, hence no additional rotations or deflections take place beyond the coronal domain of the simulation---although the CME can still decelerate, expand, and deform due to its interaction with the ambient medium. Hence, the results presented here show the importance of accurately determining CME input parameters from observations, since small spreads in predictions closer to the Sun can result in broad uncertainties at 1~au even when neglecting additional evolutionary effects in the heliosphere. For example, this is evident when considering the best fit at SolO propagated to STEREO-A, resulting in a CME ejecta arrival ${\sim}12$~h later than the best STEREO-A run and in magnetic field magnitudes approximately twice as high. When considering predictions at PSP and STEREO-A, on the other hand, we found that the same ensemble run produces the best fit at both spacecraft. The results shown here indicate that there may be a threshold (in terms of radial and angular distance) to the usefulness of inner-probe observations for 1-au predictions, beyond which correlations largely cease---in this work, at their respective CME arrival times STEREO-A is separated by 0.52~au and $12^{\circ}$ from Bepi, 0.35~au and $11^{\circ}$ from SolO, and 0.18~au and $5^{\circ}$ from PSP. Indeed, it has been shown that even in in-situ CME reconstructions (rather than forward-modelling) reconciling measurements taken at far-separated spacecraft is often not possible under the assumption of self-similar expansion \citep[e.g.,][]{weiss2021, davies2024}.

Nevertheless, we remark that in this work we have employed a simple goodness-of-fit metric based on absolute errors between modelled and observed quantities combined with timing errors. In the future, we will consider more sophisticated metrics such as dynamic time warping, which has been shown in recent works to be applicable to solar wind \citep[e.g.,][]{samara2022, kieokaew2024}, solar energetic particle \citep[e.g.,][]{palmerio2024a}, and even geomagnetic index \citep[e.g.,][]{laperre2020, maharana2024} time series. It is possible that different metrics highlight different solutions as the ``best run'', and more work is necessary in this direction to evaluate how to best benchmark CME arrival time and magnetic field configuration within a single, combined metric \citep[see][for an overview of some initial efforts on the matter]{verbeke2019}.

%% -------------------------------------- %%
%% Figure: Seed run -- southern crossings
\begin{figure*}
\centering
\includegraphics[width=.95\linewidth]{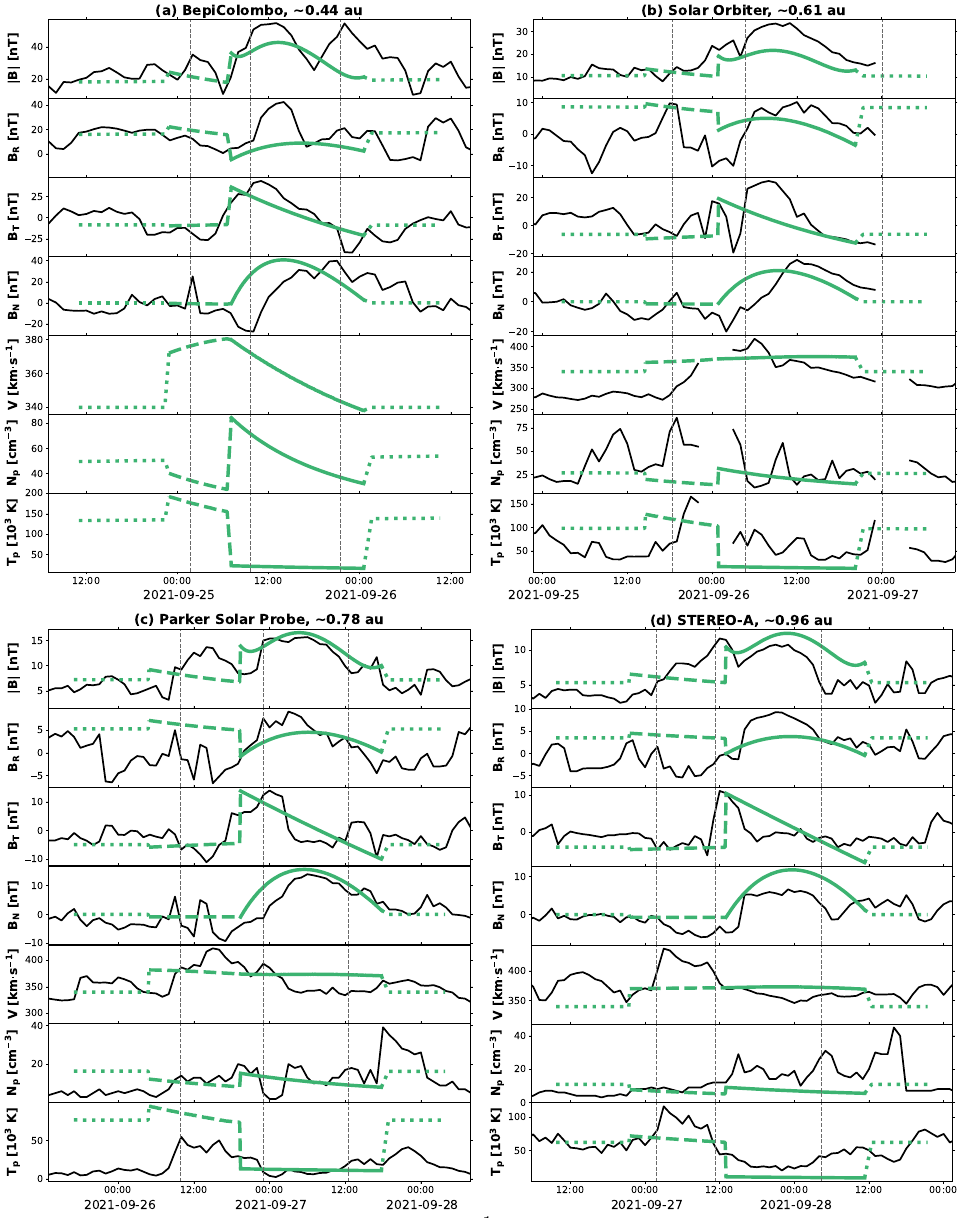}
\caption{Overview of the OSPREI seed simulation run results with latitudinally-mirrored (with respect to the CME nose) spacecraft crossings shown against in-situ measurements of the 23 September 2021 CME at (a) Bepi, (b) SolO, (c) PSP, and (d) STEREO-A. All panels and parameters are shown in the same format as Figure~\ref{fig:seed_insitu}.}
\label{fig:seed_south}
\end{figure*}
%% -------------------------------------- %%

Finally, we noted that, even if the observed $B_{T}$ and $B_{N}$ components were somewhat well captured by the seed and ensemble member runs (Figures~\ref{fig:seed_insitu} and \ref{fig:ens_insitu}), the $B_{R}$ component was predicted to have the opposite sign for all cases. To verify whether this is indeed a result of all the simulated crossings taking place north of the CME nose (see Figure~\ref{fig:ens_evo}(b)), we extract synthetic in-situ profiles from the seed run by mirroring each spacecraft's position in latitude with respect to the apex (i.e., by considering the corresponding ``minus'' Y-coordinates in Figure~\ref{fig:ens_evo}(b)), as shown in Figure~\ref{fig:seed_south}. Indeed, all $B_{R}$ predictions inside the ejecta are significantly improved, indicating that the CME may have been crossed, in reality, south of its nose by all four probes. It is possible that the ForeCAT deflections in the corona were underestimated by OSPREI (cf.\ the nose latitude of only $-8^{\circ}$ estimated via the GCS reconstructions in Section~\ref{subsec:corona} against the seed latitude of $-25^{\circ}$ in Section~\ref{subsec:seed}), and/or that the CME further deflected northwards during its interplanetary propagation \citep[e.g.,][]{isavnin2014}, where we have instead assumed constant trajectory. It is worth noting that OSPREI could also be run in ``interplanetary mode'' only, where CME parameters in the outer corona (estimated, e.g., from GCS reconstructions) are propagated directly with ANTEATR+FIDO---thus, bypassing the ForeCAT portion of the modelling suite. In the case of the event studied here, such a run (not shown) yields $B_{R}$ components that are still negative but significantly closer to zero than in Figure~\ref{fig:seed_insitu}, further suggesting that the CME might have continued its northwards deflection after leaving the solar corona.

%% -------------------------------------- %%
%  CONCLUSIONS
%% -------------------------------------- %%

\section{Summary and conclusions} \label{sec:concl}

In this work, we have taken advantage of an exceptional clustering of spacecraft between ${\sim}0.4$ and ${\sim}1$~au to perform analysis and modelling of the 23 September 2021 CME from its eruption up to its arrival at ${\sim}1$~au. To our knowledge, this is the first report of an event being observed in situ by four well-radially-separated probes inside 1~au, providing a critical opportunity to evaluate CME evolution across the different locations and to provide additional validation for multi-spacecraft CME propagation modelling. Overall, the picture that emerges from the synthesis of remote-sensing and in-situ observations of the 23 September 2021 CME is that of a slow, moderate-sized event that propagated through a slow ambient wind and that largely maintained its magnetic orientation as estimated from solar data. Nevertheless, the in-situ profiles at the four probes, whilst following similar trends, are characterised by several prominent differences---especially in the magnetic field magnitude and in a number of discontinuities present only at a subset of spacecraft (see Figure~\ref{fig:mag_stretch}). Given that strong interactions with a structured solar wind are not expected to have taken place in this particular event \editone{(since, as mentioned in Section~\ref{subsec:discobs}, the CME is embedded in the slow solar wind at all locations for which plasma data are available)}, it is possible that distortions over the full CME body are a remnant of the intrinsic complexity of the CME eruption dynamics (see Section~\ref{subsec:sun}) and coronal evolution (see Section~\ref{subsec:corona}), which involve a multipolar source region and multiple eruptions close in time. For example, \citet{bothmer2017} suggested that kinks in the near-Sun flux rope configuration can propagate through its interplanetary evolution, \editone{resulting in a CME body characterised by local deviations from the global structure that may even appear to feature different orientations from one in-situ encounter to the next}.

Modelling of the 23 September 2021 CME with OSPREI showed \editone{that} the seed run does an adequate job at predicting the \editone{various arrival times (which are also consistent with WSA--Enlil+Cone results, see Appendix~\ref{app:enlil}) as well as the} multi-spacecraft in-situ profiles at the large scale in a hindcast fashion (see Figure~\ref{fig:seed_insitu}), but is naturally not capable to reproduce the smaller-scale variability encountered in the in-situ measurements. In this sense, Sun-to-1~au MHD modelling \citep[e.g.,][]{jin2017, torok2018} is expected to be better-suited to capture the complex evolution of CME magnetic fields during interplanetary propagation. We noted that the seed run predicted an opposite sign of $B_{R}$ with respect to spacecraft observations and showed that ``mirroring'' the encounters south of the CME nose yields a better match (see Figure~\ref{fig:seed_south}), suggesting that the CME deflected further northwards than estimated in our simulations. This exercise highlights the importance of leveraging modelling results to further interpret observational data, where discrepancies in the \editone{compared} time series can be used to extrapolate and draw conclusions as to the evolution dynamics of a CME. Ensemble modelling with OSPREI revealed that the ``best-fit'' runs across the different spacecraft tend to diverge with heliocentric distance and/or angular separation, indicating that using measurements at an inner probe to constrain predictions at an outer probe is a reasonable approach as long as the two locations are not ``too far apart''. In practical terms, given that PSP (0.78~au) and STEREO-A (0.96~au) were characterised by the same ensemble member resulting in the best-fit run, it appears reasonable to assume that sub-au probes around Venus's orbit might be an optimal choice to constrain 1~au predictions whilst allowing for enough leading time \citep[e.g.,][]{szabo2023}.

Finally, analysis and modelling of the 23 September 2021 CME has been possible due to a fortuitous relative configuration of four probes inside 1~au: Bepi, SolO, PSP, and STEREO-A. Multi-spacecraft CME encounters that involve more than two probes are understandably rare and are often rather complex to interpret, since it becomes increasingly difficult to attribute differences in the measurements to heliospheric evolution (in time) and/or to local distortions (across the CME body). A dedicated constellation with well-defined spatial and angular separations is expected to provide improvements towards resolving such ambiguities \citep[see][for a detailed numerical study on the amount of probes necessary to fully characterise CME complexity]{scolini2023}. Nevertheless, as discussed in \citet{palmerio2023a}, this study represents yet another proof of the importance of multi-spacecraft measurements and of taking advantage of as many data sets as possible, including those from planetary missions (as was the case for Bepi in this work), to bring further insights into the varied aspects of CME evolution in the heliosphere.

%% -------------------------------------- %%
%  ACKNOWLEDGEMENTS
%% -------------------------------------- %%

\section*{Acknowledgements}

EP, CK, NAH, and WY are supported by NASA's Parker Solar Probe Guest Investigator (PSP-GI; grant no.\ 80NSSC22K0349) programme.
EP and PR acknowledge NSF's Prediction of and Resilience against Extreme Events (PREEVENTS; grant no.\ ICER‐1854790) as well as NASA's Living With a Star (LWS; grant no.\ 80NSSC24K1108) programmes.
BJL acknowledges support from NASA 80NSSC21K0731, 80NSSC21K1325, and 80NSSC22K0674, as well as NSF AGS 2147399.
VEL is supported by the National Science Foundation Graduate Research Fellowship under grant no.\ 2235201.
BSC acknowledges support from the UK-STFC Ernest Rutherford fellowship ST/V004115/1 and the BepiColombo guest investigator grant ST/Y000439/1.

This study has received funding from the European Unions Horizon 2020 research and innovation programme under grant agreement no.\ 101004159 (SERPENTINE; \href{www.serpentine-h2020.eu}{www.serpentine-h2020.eu}). Views and opinions expressed are, however, those of the authors only and do not necessarily reflect those of the European Union or the European Research Council Executive Agency. Neither the European Union nor the granting authority can be held responsible for them.
The authors thank NASA's Community Coordinated Modeling Center (CCMC; \href{https://ccmc.gsfc.nasa.gov}{https://ccmc.gsfc.nasa.gov}) for supporting the WSA--Enlil+Cone simulation efforts presented in this work. The WSA model was developed by C.~N.~Arge (currently at NASA Goddard Space Flight Center) and the Enlil model was developed by D.~Odstrcil (currently at George Mason University).

%% -------------------------------------- %%
%  DATA AVAILABILITY STATEMENT
%% -------------------------------------- %%

\section*{Data availability}

Remote-sensing \editone{(EUV and white-light)} data from SDO, SOHO, and STEREO are openly available at the Virtual Solar Observatory (VSO; \href{https://sdac.virtualsolar.org/}{https://sdac.virtualsolar.org}), \editone{whilst full-Sun magnetograph maps from SDO can be found at the Joint Science Operations Center (JSOC; \href{http://jsoc.stanford.edu/}{http://jsoc.stanford.edu})}. These data were visualised, processed, and analysed trough SunPy \citep{sunpy2020}, IDL SolarSoft \citep{freeland1998}, and the ESA JHelioviewer software \citep{muller2017}. 
Bepi data from the mission's cruise phase will be released to the public in the future. SolO, PSP, and STEREO-A data can be found at NASA's Coordinated Data Analysis Web (CDAWeb; \href{https://cdaweb.gsfc.nasa.gov}{https://cdaweb.gsfc.nasa.gov}) database.
The OSPREI modelling suite is entirely available online and can be found at \href{https://github.com/ckay314/OSPREI}{https://github.com/ckay314/OSPREI}.
Finally, the WSA--Enlil+Cone simulation run employed in this work can be accessed online at \href{https://ccmc.gsfc.nasa.gov/ungrouped/SH/Helio_main.php}{https://ccmc.gsfc.nasa.gov/ungrouped/SH/Helio\_main.php} (run id: \textsl{Erika\_Palmerio\_072624\_SH\_1}).

%% -------------------------------------- %%
%  BIBLIOGRAPHY
%% -------------------------------------- %%

\bibliographystyle{mnras}
\bibliography{bibliography} 

%% -------------------------------------- %%
%  APPENDIX
%% -------------------------------------- %%

\appendix

\section{CME propagation with WSA--Enlil} \label{app:enlil}

To verify that the expected in-situ impacts from the 23 September 2021 event are reasonable in terms of arrival times at the different locations, we model the inner heliospheric propagation of the CME using the coupled WSA--Enlil model. WSA operates in the so-called coronal domain of the simulation and employs magnetic field maps of the solar photosphere to generate the ambient conditions in the range 1--21.5\,$R_{\odot}$. Enlil operates in the so-called heliospheric domain of the simulation and uses WSA outputs at 21.5\,$R_{\odot}$ (or 0.1~au) to model the solar wind and interplanetary magnetic field up to a user-defined heliocentric distance---in this work, we set our outer boundary to 1.1~au. Within this framework, CMEs can be inserted at the interface between the WSA and Enlil domains, i.e.\ 0.1~au, corresponding to the outer corona. The employed CME ejecta morphology consists of a tilted ellipsoid \citep[see][]{mays2015} and lacks an internal magnetic field---we shall refer to this set up as Enlil+Cone. A CME ejecta described as a hydrodynamic pulse is not appropriate for modelling and reproducing its magnetic field configuration; nevertheless, the WSA--Enlil+Cone framework has been shown to be adequate for evaluating multi-spacecraft CME arrival times \citep{odstrcil2023}.

The photospheric maps that we use as input for WSA are daily-updated zero-point-corrected synoptic magnetograms from the National Solar Observatory (NSO) Global Oscillation Network Group \citep[GONG;][]{harvey1996}. The input parameters for the cone CME are taken directly from the GCS reconstructions presented in Section~\ref{subsec:corona} and Figure~\ref{fig:obs_wl}, and the formulas to convert GCS dimensional parameters into semi-minor and semi-major axes of the ellipsoidal CME cross-section can be found in \citet[][Appendix~A]{palmerio2023b}. The CME is inserted through the inner boundary of Enlil on 23 September 2021 at 17:08~UT with a propagation direction ($\theta$, $\phi$) = ($-8^{\circ}$, $-37^{\circ}$) in Stonyhurst coordinates, a tilt $\psi = 40^{\circ}$ (measured counterclockwise from the solar west direction), half-widths ($R_\mathrm{max}$, $R_\mathrm{min}$) = ($30^{\circ}$, $20^{\circ}$), and initial speed $V_{0} = 390$~km$\cdot$s$^{-1}$. An overview of the simulation results is displayed in Figure~\ref{fig:enlil} and an animation is provided in Supplementary Video~4.

%% -------------------------------------- %%
%% Figure: WSA--Enlil simulation
\begin{figure*}
\centering
\includegraphics[width=.91\linewidth]{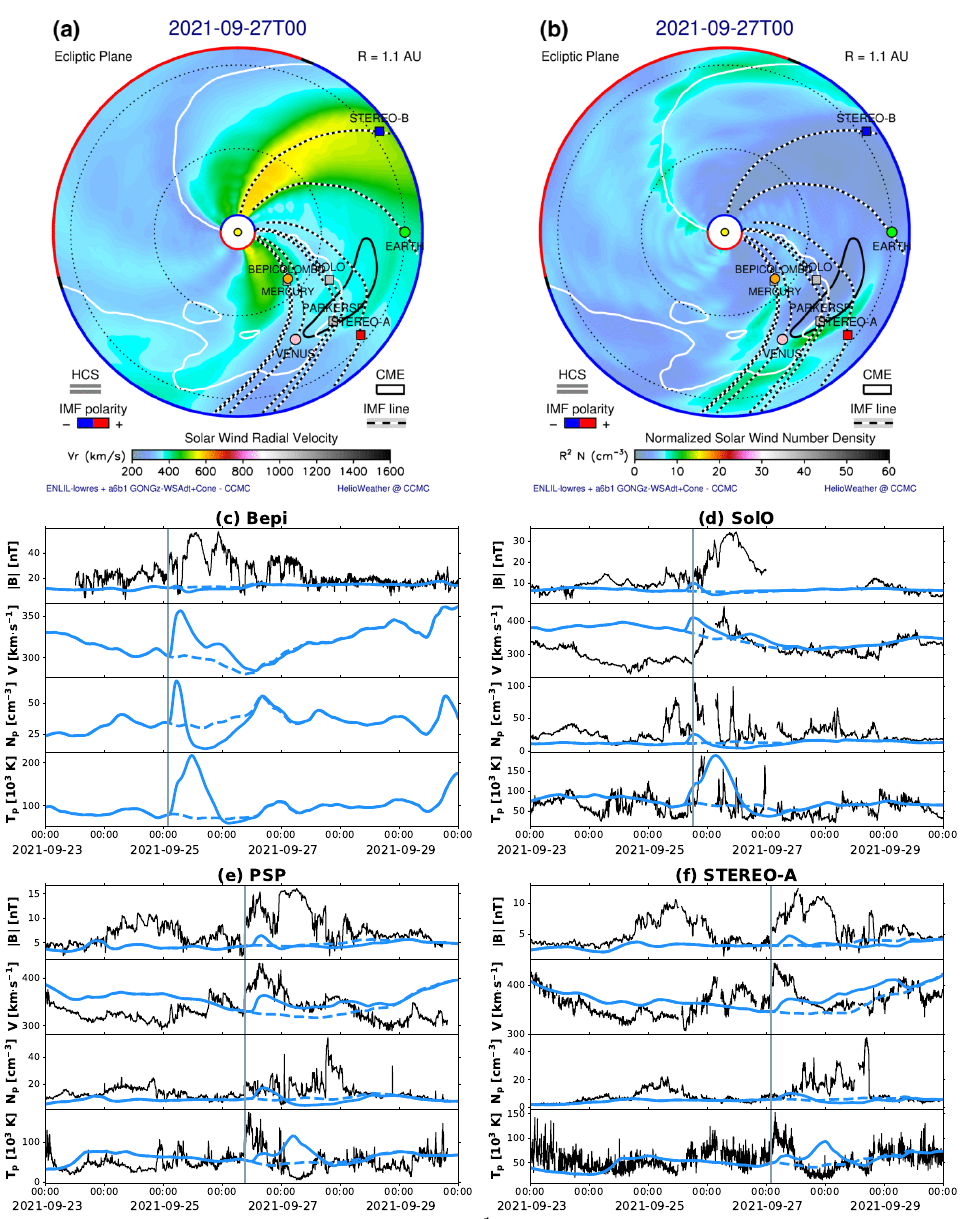}
\caption{Overview of the WSA--Enlil+Cone simulation run and comparison of the modelled results at the four spacecraft of interest. The top row shows views on the ecliptic plane of the solar wind (a) radial speed and (b) normalised number density within the simulation's heliospheric domain (0.1--1.1 au), showing the CME (outlined with a black contour) about to impact PSP. An animated version of panels (a) and (b) is featured in Supplementary Video~4. The remaining panels display WSA--Enlil simulation results presented against spacecraft measurements at (c) Bepi, (d) SolO, (e) PSP, and (f) STEREO-A. Observations are shown in black, whilst modelled time series are shown in blue---where the solid line indicates the WSA--Enlil+Cone simulation run and the dashed line provides the corresponding WSA--Enlil ambient run (without the CME). The observed interplanetary shock arrival at each location is marked with a vertical grey line.}
\label{fig:enlil}
\end{figure*}
%% -------------------------------------- %%

It is evident from the (modelled-versus-observed) time series comparisons shown in Figure~\ref{fig:enlil}(c--f) that the 23 September 2021 CME is expected to impact all four spacecraft considered in this work, i.e.\ Bepi, SolO, PSP, and STEREO-A. The simulated arrival times at each location are remarkably close to the corresponding spacecraft measurements, the modelled impacts being ${\sim}1$~h late at Bepi, ${\sim}3$~h early at SolO, ${\sim}4$~h late at PSP, and ${\sim}3$~h late at STEREO-A---all comfortably within the current CME arrival time uncertainties of ${\gtrsim}10$~h \citep{kay2024b}. Additionally, it is possible to note in Figure~\ref{fig:enlil}(a--b) and Supplementary Video~4 that the CME is expected to encounter Bepi through its very eastern flank, SolO through its nose, and PSP as well as STEREO-A at intermediate distances from the apex. Overall, the WSA--Enlil+Cone simulation showcased here demonstrates that our assessment and interpretation of the large-scale heliospheric evolution of the 23 September 2021 event is self-consistent.

%%%%%%%%%%%%%%%%%%%%%%%%%%%%%%%%%%%%%%%%%%%%%%%%%%

% Don't change these lines
\bsp	% typesetting comment
\label{lastpage}
\end{document}